# Orbital Data Confirms Dynamic Fractal Firework Universe Having 3D-spiral Code

E. P. Savov

*Solar-Terrestrial Influences Laboratory, Bulgarian Academy of Sciences*
*Acad. G. Bonchev St., Block 3, Sofia 1113, Bulgaria*

**Abstract**

The understanding of the universe is confused by the unknown nature of about 95% of its matter, required to confine the motions of space objects in cosmic structures. The idea of this paper is that the self-similar transformations of one postulated basic matter create the fundamental dynamic fractal elements of the universe. The elements are described with equation of unifying interaction drawn into their fundamental framework. The equation transforms into the inverse square laws, the principle of uncertainty, Maxwell equations of electromagnetic field and into MOND like expression after modification of its parameters at the scales where these laws and principle are found to hold. The introduced equation is also confirmed with calculations based on the orbital data for bodies that move around the nucleus of the Milky Way Galaxy, the centers of the Earth and the Sun and orbit the huge 87 Sylvia asteroid. The puzzling constant sunward force that acts on the Pioneer 10 and 11 spacecraft, the flat rotational curves of stars far from the galactic nuclei, the nature of dark matter and the enigmatic cosmic repulsion are explained in terms of contracting and expanding 3D-spirally-faster-inward-oscillating vortexes of basic matter, called dynamic fractals. The latter create new fundamental framework, which allows qualitative and quantitative modeling by fitting of the parameters of the equation of unifying interaction to experimental data. The calculations based on the proposed dynamic fractal framework are in agreement with the considered orbital data. Many testable predictions are made and natural hazard precursors are suggested.

## 1. INTRODUCTION

The mysterious matter-antimatter asymmetry in the big bang universe beginning generates finite particles that interact at finite scales. The distance between the particles increase as the big bang universe expands. Hence the strength of the interaction between the particles will decrease rather than increase and create the observed clumpiness of matter. The origin of gravitational instabilities in the big bang created expanding domain of decreasing interaction is uncertain. The pattern in which matter convenes in cosmic structures is controversial. There is also no way to build the observed dense populations of stars and clusters of galaxies without some cosmic repulsion working at stellar and galactic scales that prevents these space objects from merging. The pattern in which matter piles up into the fundamental elements of the universe has to be revealed. Then we will know how bodies attract and repel at different scales, built from the found fundamental elements.

Nature is described with what we see as space, time, bodies, waves and their properties like mass and electric charge. What creates what we see is uncertain. We miss the underlying structure that creates the images in one's mind. This makes the current understanding of the universe complex and incomplete. It is dramatically simplified with Mandelbrot's expanded notion of fractal, suggesting invariance under certain class of transforms for a set be fractal – made of self-similar elements [1]. Everything is interaction because it indicates its existence through certain interaction [2]. The laws of the all-building unifying interaction are scale independent and hence observer invariant. The scale dependence of the classical and quantum theories is likely to originate in the process of observation. Otherwise observer has to occupy a very special place - somewhere between the deterministic and continuous classical and the probabilistic and discrete quantum realities. It is much more likely that these two realities are generated in the process of perception of one poorly understood unified reality. A set of scale independent laws of unifying interaction will create dynamic fractal structure of reality. This 3D-spirally-faster-inward-moving structure of nature was found from investigation of the solar-wind magnetosphere interaction [3]. The found pattern of unifying interaction takes us into the structure of reality just by its self-reproduction in different scales. The understanding of nature depends on discovery of the fundamental framework that corresponds to its all-explaining structure.

The idea is that the universe is made of self-similar transformations of one postulated basic matter. The logic is that these transforms eject fundamental elements from the 3D-spirally oscillating insides of similar larger ones. The orbital data for Sun's motion around the Galactic nucleus, the solar system planets, Earth's satellite and the moonlets of 87 Sylvia asteroid are considered. The data confirm equation of dynamic fractal unifying interaction derived from a revealed self-similar pattern of this interaction that suggests dynamic fractal firework universe having 3D-spiral code.

It will be shown that rather from occupying a special place in one singularity plagued and incomplete big bang universe, we live in one all-building dynamic fractal structure. Its multi-scale finite elements create the discrete structure of reality. The ratio between the scales of these dynamic elements and the scales of the interaction that builds the process of observation creates what we see as space, time, bodies, continuous macro and quantum micro world. Everything is self-defined and hence self-similarly evolving interaction, i.e. basic matter. The found pattern in which it remains always finite, defines the three dimensions of the bodies and accounts for the complete and self-consistent picture of the dynamic fractal firework universe. To understand the firework universe one should forget the current knowledge and follow only its intrinsic logic. It shows how smaller fundamental elements originate from the revealed structure of similar finite larger ones. These elements are described with equation of unifying interaction, which is confirmed with orbital data, obtaining of the inverse square laws and the uncertainty principle.

It will be shown that one postulated basic matter piles up into 3D-spiral patterns, i.e. sources of interaction that account for the general picture of the universe and many poorly understood observations. The complete and self-consistent dynamic fractal firework universe shows what comes from what without starting from a singularity and ending in uncertainty as the incomplete picture of the big bang universe does.

## 2. DYNAMIC FRACTAL HAVING 3D-SPIRAL CODE

The discovery of galaxies, clusters of galaxies and heavy elements in the most distant observable universe [e.g. 4-6] suggests that they are older than the time since the big bang, allowed for their formation. The big bang universe had recurrent problems of age, which require considering of alternatives based on proper qualitative reasoning rather than numbers [7]. The real universe unfolding is essential for progress in all scientific fields because the origin of chemical elements and space bodies creates the fundamental framework for understanding of everything. The understanding of the universe is a matter of finding out how matter piles up into space bodies, atoms and whatever we see. The pattern of one postulated basic matter piling will be considered. It happens in a way that creates dynamic fractal unifying interaction, which accounts for the observed similarity between the near and most distant cosmos, the fractal dimensions of galaxies distributions and the ratio between the masses of the Earth and the Sun [2, 8]. The formula of the unifying force converges to the essential laws of classical and quantum theories, when considered at the scales of observation [2, 8]. The all-explaining basic matter distribution is discovered dynamic fractal pattern of unifying interaction. It creates new fundamental framework for qualitative and quantitative modeling of phenomena [2, 8]. In the found self-similarly evolving, fractal like, firework universe the properties of one element considered at its scales account for all [2]. This dynamic fractal element is shown in Fig. 1.

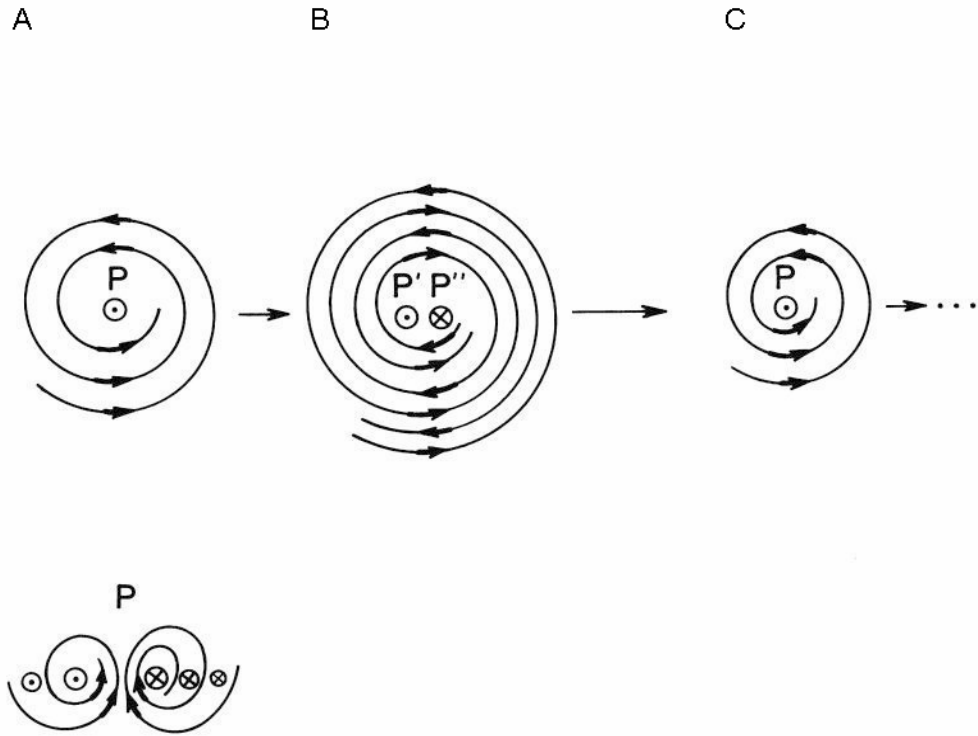

**Fig. 1.** Self-similarly evolving, 3D-spirally-faster-inward contracting and expanding, vortex of basic matter P → P' and P" → P …, called dynamic fractal of unifying interaction, creates universal distance $r_i$ (scale) from its core. The basic matter moves with velocity $v(r_i)$ 3D-spirally-faster-inward as shown by the inward increase of arrows in the equatorial (above) and meridian (below) cross-sections (A). Afterwards it bounces back and starts to oscillate 3D-spirally-faster-inward (the meridian cross-section is not shown in (B) and (C)). Thus it remains always finite, i.e. singularity free, in the smallest number of dimensions – 3D. Finally it collapses on its source (C). It oscillates 3D-spirally-faster-inward, accumulates its environment and ejects smaller similar ones that do the same (Fig. 3). *Thus it creates the finite, curved and discrete, self-defined and self-similar sources of all-building unifying interaction in one self-consistent and complete picture of firework universe that self-similarly evolves through all annihilating cycles (Figs. 1-5).* The universe collapses on its source (C and A) and is similarly born again (B), governed by similar laws of physics. The strongly prevailing faster inward 3D-spiral vortex of the basic matter crushes its environment and so shows antimatter properties. (Figure from [9])

The analysis of the dynamic fractal shown in Fig. 1 is equivalent to the analysis of the whole dynamic fractal structure due to its self-similar evolving. It is postulated to be made of one basic matter.

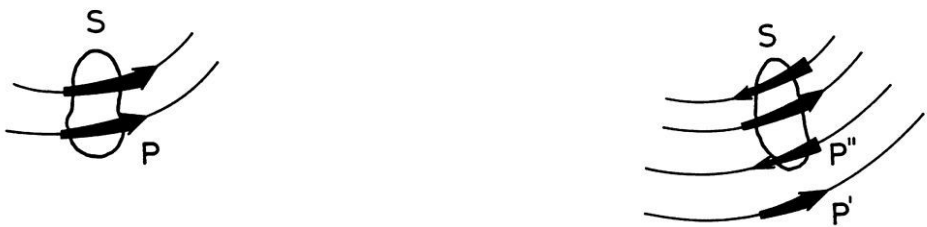

**Fig. 2.** This is a magnification of the 3D-spiral vortex of basic matter, i.e. the fundamental dynamic fractal element of unifying interaction described in Fig. 1. The basic matter that builds it moves 3D-spirally faster inward as shown by the inward increasing arrows. Thus it creates non-zero flux through every surface S and so indicates its existence. The 3D-spirally-faster-inward-moving basic matter (P and P') creates initial universal attraction that brings the dynamic fractal elements together. This accounts for the observed clumpiness of matter at human and cosmic scales. The secondary 3D-spiral-faster-outward motion P" (Fig. 1B)

creates universal repulsion that keeps the cores of the dynamic fractal elements apart. We see only the cores of the dynamic fractals, i.e. 3D-spiral vortexes of basic matter, which are dense enough to emit or reflect much smaller ones, observed as light, created as shown in Fig. 3. Everywhere $\nabla \cdot \mathbf{P} \neq 0$ holds for the 3D-spiral pattern of the velocity $\mathbf{P}$ of the basic matter. Anyway it shows up at the finite accuracy of measurement as zero and non-zero flux. Then it is modeled with $\nabla \cdot \mathbf{P}=0$ or with $\nabla \cdot \mathbf{P} \neq 0$, respectively, for what are observed as magnetic or electrostatic interactions [2]. The non-zero 3D-spiral flux of basic matter through every surface creates dynamic fractal structure that has a 3D-spiral code (Figs. 1-5). The structure self-similarly evolves through all-annihilating cycles and thus confirms its universality.

    The knowledge of what comes first tells everything in terms of origin (structure). The 3D-spirally-faster-inward contracting and expanding dynamic fractal (Fig. 1) creates similar finite fundamental elements ordered along their origin – the smaller are ejected like fireworks from the cores of the larger ones and driven into their outer 3D-spiral structure as shown in Fig. 3. The discrete structure of this interaction allows the much smaller ones to leave their sources and to be seen as light. Every body expands from its source and the expansion of the structure of light along its way from the distant galaxies will create the intrinsic nature of redshifts [2].

    Every body moves around its source as shown in Fig.3. That is why we see stars moving around galactic nuclei, atoms moving around their sources – the discovered nuclei of stars and planets, which are cores of the stellar size dynamic fractals of Fig. 1B type [2]. Planets and their planetary like moons are small cooled stars. Atomic and smaller size dynamic fractals, i.e. atomic nuclei, electrons and light photons, are created from the discovered nuclei of stars and planets. This keeps the interiors of these space bodies hot. The dynamic fractal origin of stars and atomic matter is in agreement with the finding that the bulk of the sun is made from iron and heavy elements [10] because the heavy elements are likely to remain closer to their very hot sources if born as shown in Fig. 3, while the light elements move outward to create the stellar winds. The heavy elements are later annihilated in supernova events and so the observed cosmic space, filled predominantly with light elements H and He, is created.

    In the just born current firework universe the most initial dynamic fractal began to oscillate (Fig. 1A → Fig. 1B) and so ejected smaller similar ones that did the same (Fig. 3). Thus multi-scale similar sources of unifying interaction were created. The baby firework was looked as made of multi-scale bright blue stars. The smaller moved around the larger ones all moving around the hyper huge nucleus of the universe, which is the core of the most initial dynamic fractal (Figs. 4 and 5). The cooling of this firework universe created the cosmic microwave background and its non-uniform structure. The cooling of the smaller stars created the planets and their planetary like moons. The explosive beginnings and ends of the stars delivered in the cosmic space mostly light elements because the heavier elements remained closer to their sources – cores of the stellar size dynamic fractals.

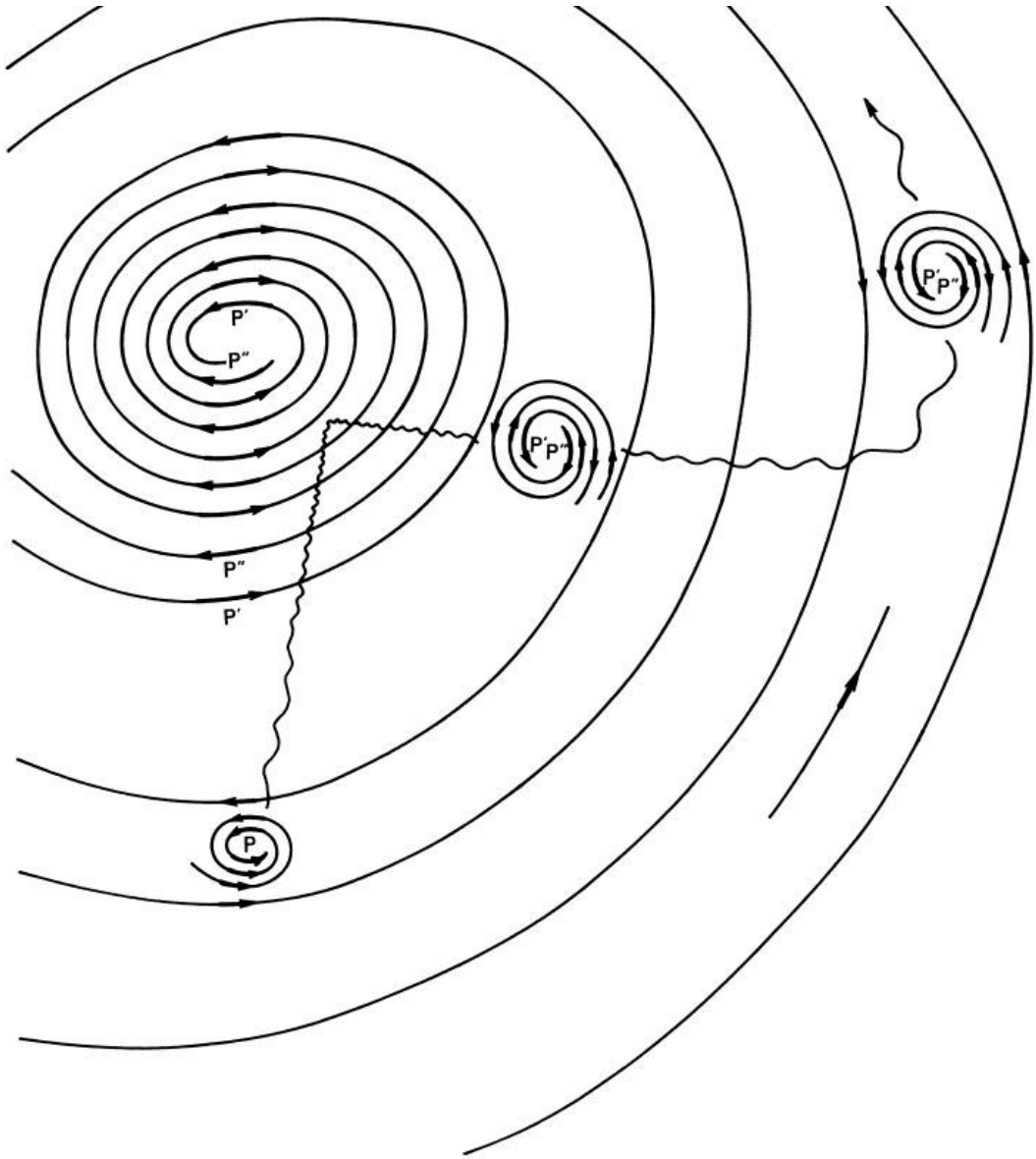

**Fig. 3.** The 3D-spirally-faster-inward contracting and expanding, oscillating, basic matter (Fig. 1B) generates similar secondary 3D-spiral vortexes (dynamic fractals) of Fig. 1A type. The secondary dynamic fractal P moves 3D-spirally faster inward and falls toward the core of the initial structure, where its density increases and it starts to oscillate (Fig. 1B). Thus it is ejected from the core of the initial dynamic fractal and driven around by its outer structure with velocity $v(r_i)$ at discrete distance $r_i$ from its core. In this way are created initial and secondary all-building dynamic fractals - 3D-spiral vortexes of basic matter, called also patterns of unifying interaction or protobodies [2]. The knowledge of origin, i.e. what comes first, tells everything [2]. The vibrating atomic size dynamic fractals eject much smaller ones, seen as light photons, which are small enough to leave the discrete structure of their 3D-spirally-faster-inward-oscillating sources. We see the oscillating insides of the 3D-spiral vortices of basic matter, i.e. dynamic fractals (Fig. 1B), which are dense enough to emit or reflect light. The faster inward motions of basic matter (shown by the larger inner arrows) create universal source ward anisotropy. (Figure from [2])

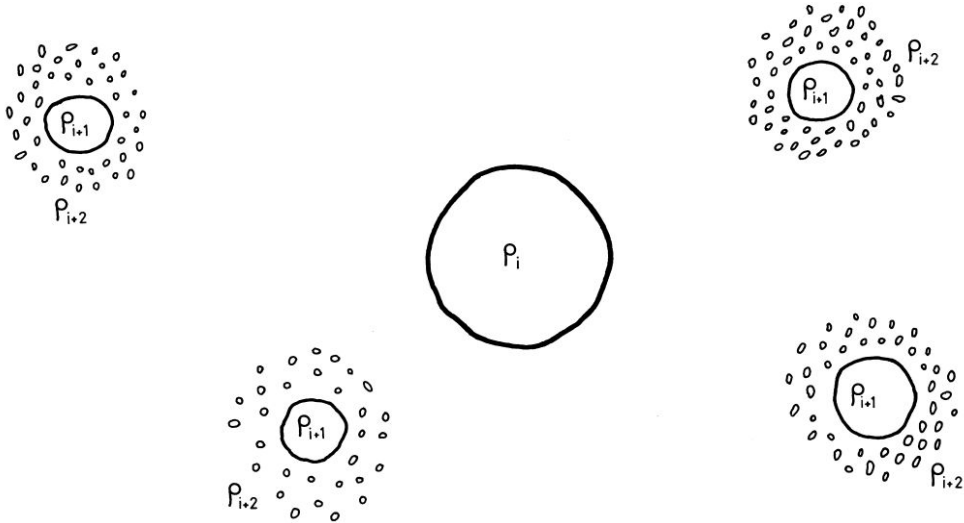

**Fig. 4.** Every body comes from the 3D-spirally-faster-inward-oscillating insides of its dynamic fractal source and moves around driven by its outer structure (Figs. 1, 3). This creates a hierarchy of origin, which is a hierarchy of scale $\rho_i$ generates $\rho_{i+1}$ that generates $\rho_{i+2}$ and so on. The hierarchy of origin begins with the most initial and hence the largest dynamic fractal of the firework universe. It 3D-spirally-faster-inward contracts, annihilates everything, and afterwards expands into a new similar firework universe governed by similar laws of physics (Figs. 1, 3). Thus it evolves self-similarly and is always self-consistently finite - singularity free, in the smallest number of dimensions - 3D that allow this to happen. The cyclic firework universe evolves through collapse of the similar previous one and consequent oscillation of the most initial dynamic fractal (Fig. 1). It ejects similar ones that do the same as it is shown in Fig. 3. The created hierarchically ordered along their origin dynamic fractals interact and build the observed clumpiness of matter at atomic and cosmic scales. The dense insides of the dynamic fractals are seen as stars, planets, moons and atoms because they emit or reflect much smaller ones, seen as light.

It exists creating a difference (distance), seen as space, time, cosmic objects, atoms and light. Then the pattern of creation of difference is the code of existence that has to be revealed. There is a fundamental asymmetry (difference) created from the 3D-spiral faster inward motion of the basic substance as it builds the dynamic fractal (Figs. 1-5). Therefore everything will show anisotropy (asymmetry) associated with its source because everything increases toward its source. The puzzling cosmic anisotropy to electromagnetic wave propagation [11] is likely to originate from the source of the observed universe, i.e. from the dynamic fractal that creates the initial space in which the waves propagate and slightly interact with the 3D-spirally-faster-inward moving basic matter that builds this seen as empty cosmic space.

The logic of the firework universe is simple. Something exists, attracts itself by moving 3D-spirally faster inward and bounces back and so on as shown in Fig. 1. This process occurs at scales of its own and creates the ordered along their revealed origin 3D-spirally-faster-inward-oscillating sources of reality. They account in the simplest and everything else excluding way for what we see [2]. The discovery of a planet under three suns [12] was predicted in terms of dynamic fractals, called also protobodies [2]. "So there should be planets around the binary and multiple stars [2, page 273]." The stellar type dynamic fractals, called stellar type protobodies [2], create what we see as coupling stars and planets as it is shown in Figs. 1-5. Hence planets are small cooled stars and can be found in binary and multiple stellar systems.

The study of the dynamic fractal shown in Fig. 1 suggests that the universe is "creation, evolution and reconfiguration of dynamic topologically equivalent basic shapes having different scales [3]". This conclusion inferred from study of the pattern of solar wind-magnetosphere interaction is also confirmed with the method of multilevel dynamical contrasting of the images. The method shows topological identity of skeletal structures having scales from $10^{-5}$ cm up to $10^{26}$ cm [13, 14]. The identical topology is preserved in self-similarly evolving shape. In other words the universe is some kind of fractal. The dynamic fractal firework universe evolves self-similarly through all-annihilating cycles and so accounts for many puzzling observations as shown in Figs. 1-5. Then this dynamic fractal, having 3D-spiral code (Fig. 1), has to be the long sought structure of reality. It will be shown that its mathematical description converges into the well tested inverse square laws at laboratory scales. The dynamic fractal structure of the universe also accounts for the mysterious sunward force that acts on the Pioneer 10 and 11 spacecraft and for the nature of dark matter.

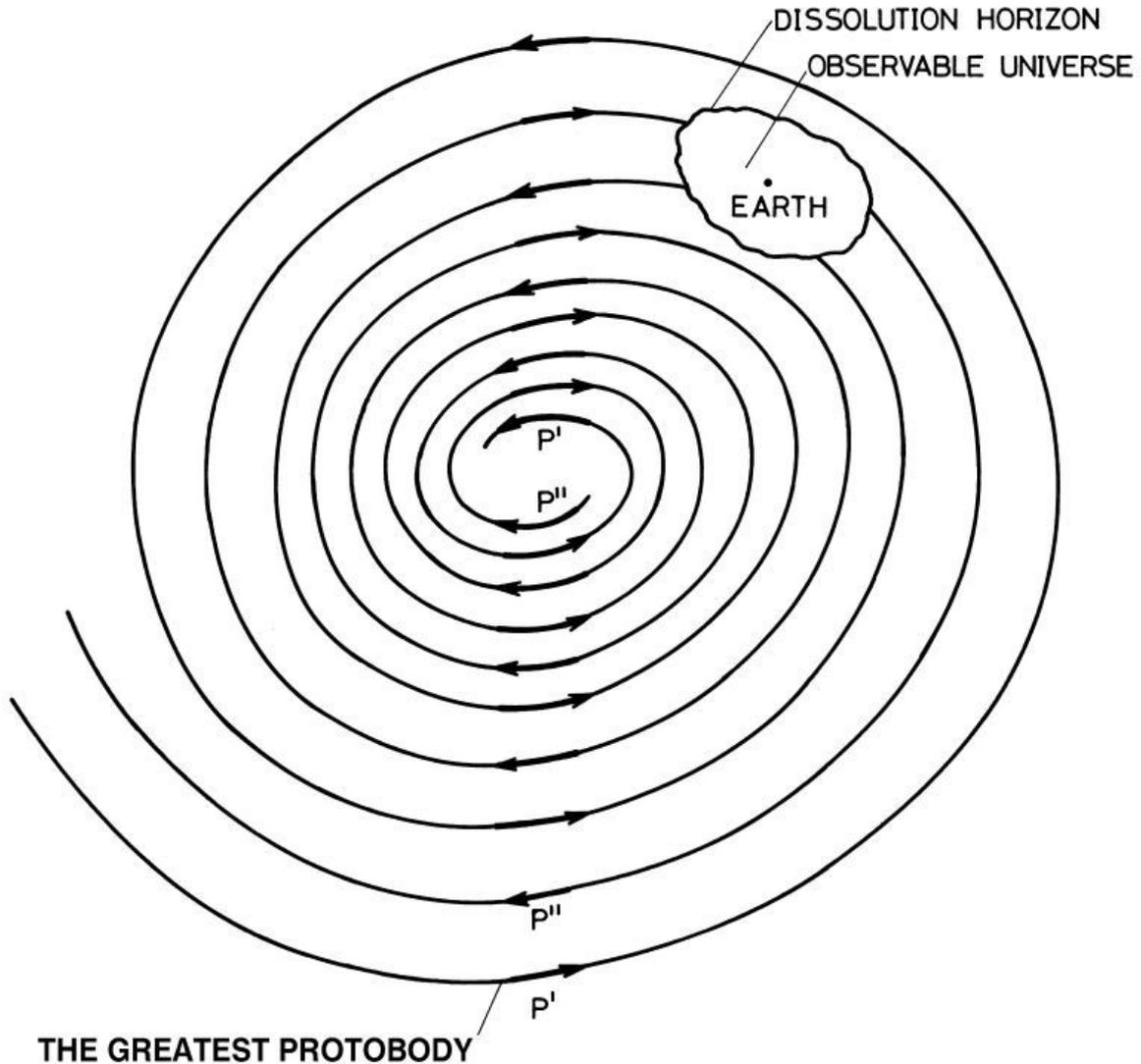

**Fig. 5.** The dynamic fractals are also called protobodies [2]. The most initial one is called The Greatest Protobody [2]. It ejects secondary ones and drives them around (Figs. 1-4). That is why similar galaxies and same chemical elements are observed at every distance from an observer who is within the dynamic fractal structure. The accessible universe is a cosmic cloud, filled with galaxies and their sources - Great Attractor [e.g. 15] type objects. The dynamic fractals of light emitted beyond the dissolution horizon, dissolve below the accuracy of observation due to decrease of their intensity. Observer in the dynamic fractal universe will see similar structures made from similar fundamental 3D-spiral vortexes of basic matter. The 3D-spiral-faster-inward piling of basic matter makes everything anisotropic inward – source ward as shown in Fig. 3. Then it is likely that the anisotropy of the distribution of galaxies will increase with redshift increase, unless the observed universe is a small fragment of the cosmos created from its source.

The dynamic fractals expand with aging as they contract and expand, accumulate their environment (Fig. 3), and create smaller ones around their cores (Figs. 3, 4). The expansion of the very small dynamic fractals seen as light is likely to create the cosmic redshift. The hierarchy of fundamental elements origin is a hierarchy of size in existing and unique cyclic firework universe that evolves self-similarly through everything else annihilating cycles as shown in Figs. 1-5. The 3D-spiral vortexes of basic matter will be described with one equation which will be confirmed with orbital data at space scales and with the inverse square laws and principle of uncertainty at atomic scales. The application of the dynamic fractals at subatomic scales is yet to be studied.

### 3. QUALITATIVE AND QUANTITATIVE ASSESSMENTS

Self-similar skeletal structures, namely cartwheels and tubules, within the range $10^{-5} – 10^{23}$ cm are found [13]. This discovery suggest that the universe is some kind of fractal [13], which is in agreement with dynamic fractal firework universe made of fundamental objects ejected from the insides of similar larger and finite ones. The observed cartwheels across the scales are created from multi-scale nuclei - dynamic fractals that make smaller ones to move around the cores of larger structures (Fig. 3). The initial elements order in chains of mutual attraction to increase their coupling with the environment. The attraction increases the interaction with environment and creates the driving force of self-organization at atomic scales, magnetic storms and substorms at magnetospheric scales and dense regions and voids at galactic scales. The expansion of the unifying interaction occurring in two dense regions keeps them apart and creates the observed less dense structure, seen as a void space between them. Fig. 1 in [16, 17] gives idea how the magnetic configuration expands during magnetic attraction. The magnetosphere 3D-spirally expands and contracts [16, 17]. Similarly the 3D-spiral swirls of basic matter expand during attraction so create dense regions and voids.

The 3D-spirally-faster-inward contracting and expanding dynamic fractal elements are discrete and show discrete (quantum) properties at their discrete scales. They can be seen as continuous motion of planets or discrete energy levels of electrons depending on their scale relative to that of the process of observation, created from discrete elements integrated in one's mind like movie frames [2]. That is why electrons abide at quantum levels and planets are found at distances from the Sun, which correspond to the Fibonacci numbers. Planets are huge dynamic fractals of Fig. 1B type, merged into even huger similar dynamic fractals whose cores are seen as stars and galactic nuclei. Every body moves into the structure of its finite dynamic fractal source. All move around the most initial dynamic fractal – called The Greatest Protobody [2]. It 3D-spirally contracts and expands and so remains always finite, i.e. not arbitrary large or small. Thus it becomes self-defined, self-similarly evolving and existing [2]. The distances between the Sun and the planets, and between the other bodies, will take discrete values corresponding to their discrete 3D-spirally-faster-inward-oscillating structures. Nothing is arbitrary in the self-defined, finite and discrete dynamic fractal structure. Every distance (difference) is self-defined by the discrete 3D-spiral dynamic fractal contractions and expansions that create it. The structure of reality is discrete made of fundamental finite dynamic fractal elements. Therefore not all real numbers have physical sense because they map on a continuous space.

The contracting and expanding, i.e. oscillating, 3D-spiral vortex of basic matter, shown in Fig. 1B, creates flux $\Phi(r)$ through a sphere having radius $r$ and center in the innermost core of the vortex. The flux $\Phi(r) = const$ defines the vortex, i.e. every dynamic fractal creates nearly constant flux of basic matter during its existence. Then the oscillating 3D-spiral vortex of basic matter (Fig. 1B) creates unifying force $F(r)$ that acts along $r$ on a body merged into it. The force is proportional to the intensity of the flux - $F(r) \sim \Phi(r)/4\pi r^2 \sim 1/r^2$ [2]. The intensity of basic matter flux creates unifying force (Fig. 1B). The force is proportional to $1/r^2$, where $r$ is the distance to the innermost core of the 3D-spiral vortex of basic matter, called dynamic fractal (element). In the collapsing structure, shown in Figures 1A and 1C, the dynamic fractal is never nearly the same because it accumulates fast new ones. This makes the distance $r$ and the sphere it defines dynamic.

Let us consider one dynamic fractal element, called simply dynamic fractal (Fig. 1A), that creates a distance $r_i$, $i=1$ from its core as it moves 3D-spirally-faster-inward. Then

$$N(r_i) = C_n[r_i]^0 = 1, \qquad (1)$$

where $N(r_i)$ is the number of dynamic fractal elements that create the distance (difference) $r_i$ and $C_n$ is a counting constant arising in the process of observation. The faster inward 3D-spiral motion of the basic matter that builds the dynamic fractal attracts new similar elements as it self-similarly evolves (Fig. 1A). Then the number of dynamic fractal elements $N(r_i)$ increases and remains finite in the smallest number of dimensions - 3D that allow this to happen (Fig. 1B). Hence for the distance $r_i$, created from the increased dynamic fractal density, it is written

$$N(r_i) = C_n[r_i]^3. \qquad (2)$$

The number of dense dynamic fractals (Fig. 1A) is proportional to the cube of the distance they create. Then the 3D-spirally-faster-inward-moving and so attracting itself dynamic fractal structure (Fig. 1A), bounces 3D-spirally outward and begins to oscillate 3D-spirally-faster-inward (Fig. 1B). Thus it remains always self-defined, self-similarly evolving, finite, discrete and unique. The frequency of its oscillation is

$$f(r_i) = 1/2T(r_i) \text{ and } T(r_i) = 2\pi r_i/v(r_i), \tag{3}$$

where $v(r_i) = const$ defines the dynamic fractal by the velocity of its elements that create the distance $r_i$, $T(r_i)$ is the time it takes for one fractal element to revolve at angle $2\pi$ into the 3D-spiral dynamic fractal structure at distance $r_i$ from its core, assuming it is nearly the same for inward and outward motions in the oscillating structure (Fig. 1B). Then the frequency of the 3D-spiral oscillation increases inward into the dynamic fractal structure as

$$f(r_i) = v(r_i)/4\pi\, r_i. \tag{4}$$

The amplitude of the universal 3D-spiral oscillation created as one element rotates at angle $2\pi$ at distance $r_i$ from the dynamic fractal core is $2r_i$. It decreases inward with frequency increase according to Equation (4). Every dynamic fractal shows $r_i \times f(r_i) = v(r_i)/4\pi = const$ at the scale $r_i$ it creates. Equations (1-4) describe the self-similarly evolving dynamic fractal. The self-similarity of this process suggests preserving of its mathematical expression. Then for the number of dynamic fractal elements it can be written

$$N(r_i) = C_n\, r_i^{d(r_i)} \tag{5}$$

where $d(r_i) \in [0, 3]$ is dynamic fractal dimension, created by $N(r_i)$ number of dynamic fractals that build the distance (universal scale) $r_i$. In the beginning of dynamic fractal evolving $d(r_i) = 0$ and afterwards it increases to $d(r)=3$ with the increase of the number of the 3D-spirally-faster-inward accumulated dynamic fractals that create the universal scale $r_i$ (Fig. 1A). The dynamic fractal dimension increases with the increase of the density of the dynamic fractals, i.e. with the increase of the number of dynamic fractals that create the distance $r_i$. Then the dynamic fractal starts to oscillate 3D-spirally-faster-inward with frequency $f(r_i)$ (Fig. 1B and Eqs. (3) and (4)). Hence from Equations (3), (4) and (5) the number of fractal elements $N_o(r_i)$ that create the universal scale $r_i$ in the oscillating structure is

$$N_o(r_i) = f(r_i)N(r_i) = [C_n v(r_i)/4\pi] \times [r_i]^{d(r_i)-1} \sim [r_i]^2. \tag{6}$$

The number $N(r_i)$ of contracted fractals (Fig. 1A) is multiplied with frequency of their oscillation to obtain the number of 3D-spirally oscillating fractals that create the distance $r_i$ as shown in Fig. 1B. The dynamic fractal starts to oscillate 3D-spirally-faster-inward and so "comes to life" as it is ejected from its similar parent structure (Fig. 3). The distance $r_i$ is made from 3D-spirally-faster-inward-oscillating dynamic fractal elements that are self-defined, finite, discrete, unique and self-similarly evolving. Figs. 1-5 show that nature is made of similar 3D-spiral oscillators, whose frequencies $f(r_i)$ increase inward, i.e. for smaller $r_i$, and the amplitudes increase outward, i.e. for larger $r_i$. The craving of nature for synchronization at far different scales [e.g. 18] is explained with its structure made of dynamic fractal 3D-spiral oscillators (Figs. 1-5) that seek to synchronize and so unfold the energy store in them. This is likely to account for the puzzling driving force behind self-organization and life.

The 3D-spiral vortex of basic matter, called dynamic fractal (Fig. 1), creates unifying force $F(r_i)$ at distance $r_i$ within its structure. The force is proportional to the dynamic fractal density $F(r_i) \sim 1/N(r_i)$, where $N(r_i)$ is the number of dynamic fractal elements building the fractal (the vortex). This number is specific for each dynamic fractal and is proportional to its resistance to acceleration $M$ called mass ($N(r_i) \sim M$). The 3D-spiral vortex of basic matter (Fig. 1) creates unifying force that drives a body around its core with acceleration $v^2(r_i)/r_i$ along $r_i$, where $v(r_i)$ is the velocity of basic matter at distance $r_i$ from the core. Then from Equations (5) and (6) the following equation of unifying interaction is obtained

$$F(r_i) = ma(r_i) = mv^2(r_i)/r_i = C_i/r_i^{d(r_i)}, \tag{7}$$

where $C_i$ is unifying force constant appearing at the scales $r_i$ of existence of the considered dynamic fractal that creates the discrete distance $r_i$, $i = 1, 2, 3, \ldots, N(r_i)$ is finite number denoting the distances $r_i$ created from the core of the discrete 3D-spiral vortex of basic matter (Fig. 1), $d(r_i) \in [0, 3]$ for 3D-spirally contracting dynamic fractal (Fig. 1A) and $d(r_i) = 2 - d_e(r_i) + d_c(r_i)$ for oscillating one (Fig. 1B), $d(r_i)$ is called dynamic fractal dimension, $d_e(r_i) \in [0, 3]$ and $d_c(r_i) \in [0, 3]$ are the dynamic fractal dimensions, respectively, associated with expansion and contraction of the 3D-spirally-faster-inward-oscillating structure, i.e. with the increase of the number of its expanding and contracting dynamic fractal elements. The oscillating dynamic fractal either expands or contracts and so $d_e(r_i) \times d_c(r_i) = 0$, $v^2(r_i)$ is the velocity of the basic matter that builds the dynamic fractal (the 3D-spiral vortex) at distance $r_i$ from its core (Fig. 1), $a(r_i)$ is the acceleration along $r_i$ created by the 3D-spirally-faster-inward motion of the basic matter and $m$ is the resistance of a body to this acceleration called mass. It can be seen from Fig.1 that $r_i$ can take only discrete values in the discrete 3D-spiral structure of the dynamic fractal. **The unifying force is discrete (quantum) due to the discreteness of the all-building 3D-spirally-faster-inward-oscillating dynamic fractal elements. The basic matter moves faster at a given distance in the larger dynamic fractal.** The mass of the 3D-spiral vortex of basic matter, i.e. dynamic fractal, is $M = C_M v^2(r_i) r_i$, where $v(r_i)$ is the velocity of basic matter at distance $r_i$ from its core and $C_M$ is a constant at the scales of the considered interaction.

The Equation of unifying interaction (7) is expressed with continuous elements that belong to finite intervals as shown in Equation (7'). The expansions and contractions of the faster-inward-oscillating 3D-spiral vortexes of basic matter described in Equation (7') account for the well known form of the physical laws and also for generation of particles, waves, atoms, planets, stars and galaxies in dynamic fractal firework universe.

$$F_u(r) = ma_u(r) = C_m(r)v_m^2(r_m)r_m v_u^2(r)/r = C(r)/r^{2 - d_e(r) + d_c(r)}, \qquad (7')$$

where $m = C_m(r)v_m^2(r_m)r_m$ is the resistance to acceleration, called mass, of a body driven by a flow of basic matter having velocity $v_u(r)$ at distance $r$ from the core of its 3D-spirally-faster-inward moving pattern that creates unifying acceleration $a_u(r) = v_u^2(r)/r$, $C_m(r)$ and $C(r)$ are constants at the finite scale $r \in [r_1, r_3]$ of the considered interaction, $d_e(r) \in [0, 3]$, $d_c(r) \in [0, 3]$ and $d_e(r) \times d_c(r) = 0$ are the dynamic fractal dimensions, respectively, of expansion and contraction of the 3D-spirally-faster-inward oscillating basic state ($d_e(r) = 0$, $d_c(r) = 0$) of the oscillating vortical source of unifying interaction, $v_m(r_m)$ is the velocity of the basic matter that builds the driven body at distance $r_m$ from its core.

The length of the intervals of expansion $d_e(r) = 1, 2, 3$, $r_{i+l} < r < r_{i+l+1}$, $l = 1, 2, 3$ decreases inward into the oscillating 3D-spiral vortex. The self-similarity of the 3D-spirally-faster-inward-oscillating basic matter allows writing for its contraction and expansion

$d_e(r) = 0$ and $d_c(r) = 0 \rightarrow F_u(r_i < r < r_{i+1}) \sim r^{-2}$, $v_u(r_i < r < r_{i+1}) \sim r^{-1/2}$

$d_e(r) = 1 \rightarrow F_u(r_{i+1} < r < r_{i+2}) \sim r^{-1}$, $v_u(r_{i+1} < r < r_{i+2}) \sim r^0$  $\qquad d_c(r) = 1 \rightarrow F_u(r_{j+1} < r < r_{j+2}) \sim r^{-3}$
$d_e(r) = 2 \rightarrow F_u(r_{i+2} < r < r_{i+3}) \sim r^0$, $v_u(r_{i+2} < r < r_{i+3}) \sim r^{1/2}$  $\qquad d_c(r) = 2 \rightarrow F_u(r_{j+2} < r < r_{j+3}) \sim r^{-4}$
$d_e(r) = 3 \rightarrow F_u(r_{i+3} < r < r_{i+4}) \sim r^1$, $v_u(r_{i+3} < r < r_{i+4}) \sim r^1$  $\qquad d_c(r) = 3 \rightarrow F_u(r_{j+3} < r < r_{j+4}) \sim r^{-5}$
$i = 1, 2, 3, ..., L$  $\qquad\qquad\qquad\qquad\qquad\qquad\qquad\qquad\qquad j = 0, 1, 2, 3, ..., K$

Here is Equation (7') applied at different scales.

**Atomic scale:**

$d_e(r) = 0$ and $d_c(r) = 0$

$F_u(r_i < r < r_{i+1}) \sim r^{-2}$, $E \sim r^{-2}$, $\nabla.E = \rho/\varepsilon$

$F_u(r) = F_E(r) = -(1/4\pi\varepsilon)q_1 q_2 / r^2$ is the Coulomb's law of electric interaction obtained from the unifying force by fitting constants, seen as electric charges $q_1$ and $q_2$ and permittivity of the medium $\varepsilon$ at the scales of the considered interaction. Hence $E \sim r^{-2}$, $\nabla.E = \rho/\varepsilon$, where **E** is called electrostatic field that shows charge density $\rho$ at atomic scales that contain its atomic scale sources.

$d_e(r) = 1$, the expansion starts from the outer elements of the dynamic fractal vortex

$F_u(r_{i+1} < r < r_{i+2}) \sim r^{-1}$, $B \sim r^{-1}$, $\nabla.B = 0$ $(r_{i+1} < r < r_{i+2}) \sim r^{-1}$

This flux of basic matter creates unifying force $F_u(r_{i+1}<r<r_{i+2}) \sim a_u(r_{i+1}<r<r_{i+2}) \sim 1/r \sim B$, which has the form of Biot and Savart's law to a point of experimentally obtained constants - $B = \mu_0 I/2\pi r \sim 1/r$, where $B$ is the magnetic field flux density at distance $r$ from a long straight conductor, $I$ is the electric current intensity and $\mu_0$ is the permeability of free space. The field $B$ describing this force will show zero divergence ($\nabla \cdot B = 0$) at the finite scales of the core of the atomic size vortex. The circulation of $B$ arises from $E$ created alignment of the atomic size basic matter vortexes. Then $\nabla \times B = \mu \sigma E + \mu \varepsilon \partial E/\partial t$, where $\mu$, $\sigma$ and $\varepsilon$ are permeability, conductivity and permittivity of the medium, respectively.

$d_e(r) = 2$

$F_u(r_{i+2}<r<r_{i+3}) \sim r^0 \sim E$, $\nabla \cdot E = 0$ because $E$ appears at the scales of $B$, which shows zero divergence. $\nabla \times E = -\partial B/\partial t$ because $E$ originates after $B$ as it increases its fractal dimension to be seen as induced electric field which can align atomic size basic matter vortexes and thus create what we see as electric current.

$d_e(r) = 3$

$F_u(r_{i+3}<r<r_{i+4}) \sim r^1$, $v_u(r_{i+3}<r<r_{i+4}) \sim r^1$ the increase of velocity creates attraction and the expanding oscillating 3D-spiral vortex begins to contract.

$d_c(r) = 1$, the contraction starts from the inner elements of the dynamic fractal vortex

$d_c(r) = 1 \rightarrow F_u(r_{j+1}<r<r_{j+2}) \sim r^{-3}$

$d_c(r) = 2$

$d_c(r) = 2 \rightarrow F_u(r_{j+2}<r<r_{j+3}) \sim r^{-4}$

the contraction deepens and the outer shells of the vortex peel off in what is seen as electromagnetic waves

$d_c(r) = 3$

$d_c(r) = 3 \rightarrow F_u(r_{j+3}<r<r_{j+4}) \sim r^{-5}$

the contraction deepens further and the vortex ejects smaller similar ones, e.g. some seen as light at atomic scales, atomic size vortexes at stellar type scales and stellar type ones at galactic ones.

> **Atomic scale:**

$d_e(r) = 0$ and $d_c(r) = 0 \rightarrow F_u(r_i<r<r_{i+1}) \sim r^{-2}$, $v_u(r_i<r<r_{i+1}) \sim r^{-1/2}$

$T(r_i<r<r_{i+1}) \sim 2\pi r/r^{-1/2} \sim r^{3/2} \rightarrow T^2(r_i<r<r_{i+1}) \sim r^3$ is the third Kepler's law

The unifying force created from the 3D-spiral motion of basic matter turns into Newton's law of gravity for stellar type, galactic nuclei and larger nuclei.

$d_e(r) = 1$, the expansion starts from outer elements of the dynamic fractal vortex

$d_e(r) = 1 \rightarrow F_u(r_{i+1}<r<r_{i+2}) \sim r^{-1}$, $v_u(r_{i+1}<r<r_{i+2}) \sim r^0 = $ const

the rotational curves of stars flatten far from cores of galactic size vortexes as they expand

$d_e(r) = 2$

$d_e(r) = 2 \rightarrow F_u(r_{i+2}<r<r_{i+3}) \sim r^0 = $const, $v_u(r_{i+2}<r<r_{i+3}) \sim r^{1/2}$

Pioneer sunward acceleration
Cartwheel galaxy formation

$d_e(r) = 3$

$d_e(r) = 3 \rightarrow F_u(r_{i+3}<r<r_{i+4}) \sim r^1$, $v_u(r_{i+3}<r<r_{i+4}) \sim r^1$

$d_c(r) = 1$, the contraction starts from the inner elements of the dynamic fractal vortex

$d_c(r) = 1 \rightarrow F_u(r_{j+1}<r<r_{j+2}) \sim r^{-3}$

$d_c(r) = 2$

$d_c(r) = 2 \rightarrow F_u(r_{j+2}<r<r_{j+3}) \sim r^{-4}$

the contraction deepens and the outer shells of the vortex peel off

$d_c(r) = 3$

$d_c(r) = 3 \rightarrow F_u(r_{j+3}<r<r_{j+4}) \sim r^{-5}$

the contraction deepens further and the vortex ejects smaller similar ones, e.g. some seen as atomic size vortexes at stellar type scales and stellar type 3D-spirally-faster-inward-oscillating vortexes at galactic scales. The recurrent contractions push the secondary vortexes outward. So galaxies, stars and planets expand before collapsing on their sources.

**< Atomic scale:**

The 3D-spirally-faster-inward-oscillating vortexes, seen as light, slowly expand like that of planets, stars and galaxies. This creates the cosmic red shift

The understanding of what comes first shows the all-explaining origin (structure) of the bodies that accounts for what we see and describe in the laws of modern physics [2]. Everything is made of finite number of finite, i.e. not arbitrary large or small, elements. Nothing is arbitrary because everything is self-defined by the interaction of its parts. The self-definiteness indicates self-similarity – a self-similarly evolving structure. The universal scale $r$ or in other words what we see as space, time and everything else collapses into the fundamental 3D-spiral basic matter vortexes build-up (Fig. 1). The collapsed universal scale appears with their oscillation (Figs. 1B, 3, 4, 5). Then the unifying force $F(r)$ increases from 0 to 3 in the collapsing structure on the way toward oscillation created $d(r) \neq 2$.

The increase in every beginning generates attraction (Fig. 1A), which turns into repulsion as it bounces from itself and thus remains always finite in the smallest number of dimensions 3D that allow it to come to being as shown in Figs. 1B and Fig. 3. The origin of the finite fundamental elements of the firework universe is described in Figs. 1-5. The secondary elements are ejected like fireworks from the insides of finite similar ones (Fig. 3). **The dynamic fractal firework universe is made of finite number of finite 3D-spiral swirls of basic matter, called dynamic fractals (Figs. 1-5).** They are described with equation of unifying interaction that can be fitted to the scales of the considered phenomena and modeled with the continuous expressions of the inverse square laws, where these laws are tested to hold. The logic of the dynamic fractal firework universe is the logic of all-building unifying interaction because this logic deals with finite fundamental elements. The flux of basic matter is everywhere different from zero and so it is finite (Fig. 2). The pattern of this flux evolves self-similarly in everything else annihilating cyclic manner that guarantees the existence of singularity free firework universe, made of finite, discrete similar, self-reproducing building blocks.

**The Equation of unifying interaction (7) holds within a finite accuracy for sets of discrete (finite) $r_i$ and $C_i$.** Then it can be modeled with continuous $r$ and $C$ for $r_i \in [r_m, r_n]$ and $C_i \sim r^\alpha$, where $i \in [m, n]$ and $\alpha \in [0, 2]$ for the force created from the 3D-spiral motion of the basic matter in the oscillating structure $\alpha = 0$. The exponent $\alpha = 1$ far from the core of oscillating structure, where it acquires 3D-spiral swirls of basic matter, i.e. dynamic fractal elements, from the initial structure at rate $\sim r^1$ and so creates constant velocity of basic matter there (Eq. (7)). For instance, one can consider the nearly constant velocities of stars, driven by the basic matter swirling around the galactic 3D-spiral vortexes, seen as spiral galaxies. The secondary structure can expand due to the direct entry of initial dynamic fractal elements in a way similar to the 3D-spiral magnetic expansion during magnetic attraction, shown in Fig. 1 in [16, 17]. Then the flux of the entering basic matter is $\sim r^2$ and so $\alpha = 2$ within the region of expansion and extra constant inward acceleration is created there, e.g. the constant sunward acceleration acting on the Pioneer spacecraft. The self-similar evolution of the dynamic fractals (Figs. 1-5) is captured into the equation of unifying interaction. This universal equation has to be drawn at the scales of the considered interactions and then its parameters must be obtained for these scales from experimental data.

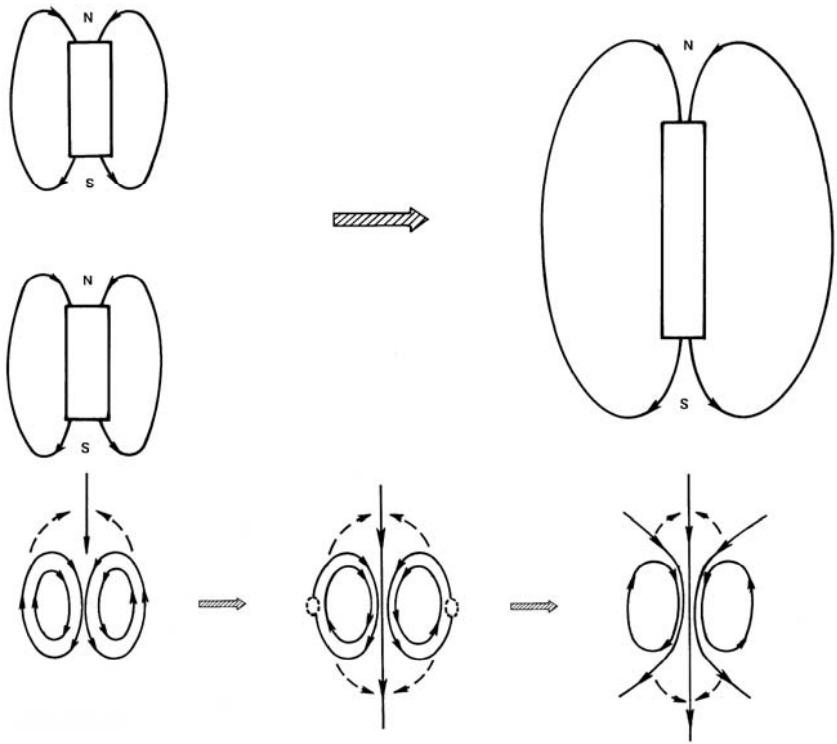

**Fig. 6.** Two sources of magnetic field face their opposite poles and attract. Then the entering external field flux is $\sim r^2$, where $r$ is the radius of a circle from the center of the magnetic configuration in its equatorial plane. The circle is perpendicular to the entering field over the northern pole of the magnetic configuration. The flux is proportional to the area of the circle ($\sim r^2$) and enters directly into the magnetic configuration through its northern polar region, and exits from the southern polar region. So the configuration expands as lines of magnetic force move 3D-spirally outward (the dashed arrows). The entering field winds up 3D-spirally (Fig. 1A) into inner region through area $(r + d)^2 - r^2 \sim r^1$, where $d$ is the width of the band in magnetic equatorial plane through which the field winds up and $d/r \ll 1$. The self-similarity of the 3D-spiral structure of interaction (Fig. 1) allows appearance of alternating regions of, respectively, direct and indirect flux entry $\sim r^2$ and $\sim r^1$. (Figure from [17], adapted from Fig. 1 of [16])

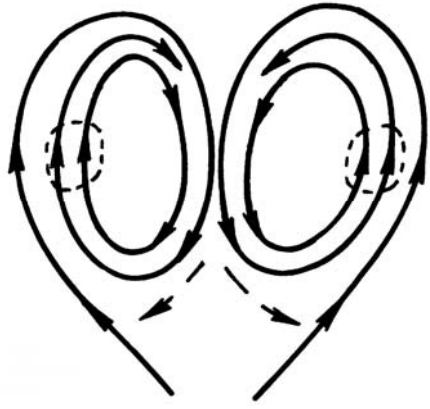

**Fig. 7.** Two sources of magnetic field face their same poles and repel. Then the external field flux enters indirectly through a band $d << r$ in the equatorial plane of the field configuration. So the entry flux is proportional to $r$ ($\sim r$), where $r$ is the distance from the center of magnetic configuration in its equatorial plane to the region of external flux entry having width $d$. The field enters indirectly into the northern polar region of the magnetic configuration and winds up 3D spirally into it. The self-similarly evolving 3D-spirally contracting and expanding structure of magnetic interaction is equivalent to that of the dynamic fractal shown in Fig. 1. This spiral magnetic field **B** structure can show zero flux through every closed surface at the finite, i.e. not arbitrary high, accuracy of magnetic interaction measurement. That is why the spiral magnetic structure allows $div\mathbf{B} = 0$ description and the topological crisis associated with it [17]. (Figure from [17], adapted from Fig. 1 of [16])

The mass of the basic matter driven around the oscillating vortical structure is $M \sim S = \pi(r + d)^2 - \pi r^2 = 2\pi r d + \pi d^2$, where $S$ is the area of indirect external flux entry between circles having radii $r + d$ and $r$, which lay in the equatorial plane of the 3D-spiral vortex and are centered in its core (Fig. 7 and Fig. 1). Into the Keplerian region of the vortex (Fig. 1B) the width of the area of external basic matter entry is $d \sim 0$, i.e. bodies move within nearly close, 3D-spirally-faster-inward-oscillating structure whose motion drives them and obeys the Kepler's laws. Let $d = const$ and $d = l$, where $l/r << 1$ is the upper scale of the secondary bodies, e.g. stars, driven around by the basic matter that winds into the initial (parent) galactic size 3D-spiral vortex. Then from the Equation of unifying interaction (7) the rotational curve becomes flat outward from the inner Keplerain domain because $v^2(r) \sim 2\pi d + \pi d^2/r^2$ and $v(r) \approx const$, where $d^2/r^2 << 1$. Thus the basic matter vortexes drive around bodies with nearly constant velocities. The increase of the inward 3D-spiral vortex will show slight indications for dark matter also in the inner Kepler's domain of the 3D-spiral vortex of basic matter. The 3D-spirally-faster-inward-osicilating galactic size vortexes of basic matter, seen as spiral galaxies, are likely to be more opened. Hence they will wind up around a thicker basic matter whirls that drive stars around with nearly constant velocities. The opening of the oscillating 3D-spiral vortex $d \neq 0$ creates extra inward force produced from basic matter entry. This can account for the sunward acceleration acting on the Pioneer spacecraft and for the flattening of rotational curves of stars far from the galactic nuclei. The stellar type 3D-spiral vortexes of stars, planets and planetary like moons, whose cores are in the centers of these spaces and are sources of their atomic shells, will show some opening, i.e. departure from the Kepler's laws. There should be a research project with multiple spacecraft around the Earth and other space bodies to test this prediction.

The magnetic configuration 3D-spirally expands (contracts) during magnetic attraction (repulsion) as shown in Figs. 6 and 7. More precisely the direct entry of the external field creates expansion with higher dynamic fractal dimension than the indirect one, which looks like relative contraction in this frame of reference. These figures reveal the 3D-spiral structure of the seen as fundamental magnetic interaction. This structure is similar to that of the dynamic fractal described in Fig.1. 3D-spiral vortex of basic matter that interacts at the scales of what we see as magnetic interaction is likely to create the magnetic interaction. It can account for the direct reversals of the geomagnetic field as it discretely evolves. The creation and deepening of magnetic storms requires threshold values of the interplanetary magnetic field [19]. Hence the expansions and contractions of the dynamic fractals, described in Figs. 1, 6 and 7, will require threshold of external basic matter flux entry. The galaxies can become disturbed in way similar to the Earth's magnetosphere that is by contraction and expansions of their galactic scale 3D-spiral vortexes of basic matter. These contractions and expansions can be caused from initial (parent) vortexes that couple with the galactic ones if the initial vortexes are above certain thresholds. The contraction and expansion of the 3D-spiral vortex of basic matter create and perturb the smaller ones in its structure. So we see perturbed galaxies and/or atoms depending on the thresholds of this process of universal perturbation, created from accumulation of external structure and its release to smaller scales. It is interesting to notice for the indirect perturbation, shown in Fig. 7, that $N(r) \sim M \sim r$, where $N(r)$ is the number of 3D-spiral vortexes of basic matter, i.e. dynamic fractals that wind up on the oscillating structure and create extra inward force accounted for as dark matter. The direct (indirect) basic matter entry into the 3D-spirally oscillating vortex leads to expansion (contraction) for certain thresholds for the entry flux (Figs. 6 and 7). The increase of the mass of the vortex for direct (indirect) entry is $M \sim r^2$ ($M \sim r$).

The Equation of unifying interaction (7) in the oscillating dynamic fractal structure ($d(r_i) = 2$) is a general inverse square law. The changing of its parameter $C_i$ to $C_i = Gm_1m_2 = const$ transforms it into Newton's law of gravity or into Coulomb's law of electric interaction for $C_i = -(1/4\pi\varepsilon)q_1q_2 = const$. Here $G$ is the gravitational constant and $m_1$, $m_2$ are masses of coupling two spheres, which are also constants during the considered interaction. Similarly, $q_1$, $q_2$ are the electric charges in which the discrete unifying interaction is seen to quantify at atomic scales and $\varepsilon$ is the absolute permittivity of the medium. We should give conventional signs of the electric charges of electrons (-), protons (+) and heavier nuclei (+). So we model their attraction and repulsion, resulting respectively, from the far different and similar sizes of these dynamic fractals. Electrons repel electrons. Similarly, but on larger scales protons repel protons and heavier atomic nuclei due to the outward 3D-spiral swirl of basic matter building their structure (Fig. 1B). The dynamic fractals seen as electrons are attracted by about $2 \times 10^3$ to $5 \times 10^5$ times larger dynamic fractals, which are seen as that times more massive chemical elements.

The Equation of unifying interaction (7), written in the dynamic fractal fundamental framework (Figs. 1-5), gives

$$F(r_i) = const. \tag{8}$$

Equation (8) means that the dynamic fractal (Fig. 1B) remains nearly the same at the scales of its existence and so creates nearly constant unifying force $F(r_i)$ in its structure at distance $r_i$ from its core. Equations (7) and (8) allow writing for oscillating dynamic fractals

$$F(r_i) = ma(r_i) = mv^2(r_i)/r_i = C_i/r_i^2 = const \text{ and hence}$$

$$v^2(r_i) \times r_i = const, v^2(r_i) = const, r_i = const \text{ and } \Delta p_x \times \Delta x = const \tag{9}$$

where $v^2(r_i) = const$ and $r_i = const$ define the considered dynamic fractal by the velocity $v(r_i)$ of its basic matter at distance $r_i$ from its core that creates the unifying force $F(r_i)$, $m$ - called mass is resistance to acceleration of a secondary dynamic fractal, e.g. one seen as a body, merged into the considered 3D-spirally-faster-inward-moving dynamic fractal structure of the initial (larger) one that creates the unifying force $F(r_i)$ and $a(r_i)$ is the acceleration along $r_i$, created from the 3D-spiral contractions and expansions of the considered dynamic fractal (Fig. 1B). Equation (9) has the sense of Heisenberg's uncertainty principle, where $\Delta p_x = 2mv(r_i)$ is the change of the impulse of the particle, e.g. electron, along coordinate $x$ as it rotates around the core of the much larger dynamic fractal, e.g. one seen as proton, $\Delta x \sim r_i$ is the accuracy of measurement of the coordinate of the particle. Equation (9) tells that the increase of the accuracy in the coordinate $\Delta x$ means increase of the number density of the dynamic fractals that create $\Delta x \sim r_i$. This drives around faster the electron nearer the core of the larger dynamic fractal, seen as atomic nucleus that creates the distance $r_i$. Therefore the direction of the velocity of the smaller dynamic fractal, i.e. the electron, smears. Then the coordinate of its impulse becomes uncertain. Similarly, the knowledge of the direction of the electron's velocity requires scales, much smaller than the atomic ones which determine the position.

Every body (dynamic fractal) is driven near or further from the core of the initial dynamic fractal structure (Fig. 3) and its trajectory has smaller or larger curvature ($1/r_i$). The equation of unifying interaction (Eq. (7)) shows that the dynamic fractal dimension $d(r_i)$ increases up to 3 with density increase (Fig. 1A). Afterwards the structure begins to oscillate 3D-spirally-faster-inward and its dynamic fractal dimension drops to 2 although its dense insides are seen as 3D-bodies that can emit or reflect much smaller ones seen as light (Fig. 1-3). The application of dynamic fractals at subatomic scales is a subject of future research. It can be noticed that the collision of two atomic nuclei of Fig. 1B type can lead to generation of antimatter particles of Fig. 1A type, showing increased interaction cross-section and taking part in destructive interactions due to lack of outward 3D-spiral basic matter swirl [2].

The velocity of stars far from the nuclei of spiral galaxies is nearly constant implying existence of some invisible dark matter $M$ that increases linearly with the distance $r$ from the nuclei [e.g. 20]. The constant velocity $v(r)$ at distance $r$ from the galactic nuclei results from $m_s[v(r)]^2/r = F(r) = C/r^{d(r)}$ (Eq. (7)), where $m_s$ is the resistance to acceleration, i.e. the mass, of the dynamic fractal element, seen as a star that moves at distance $r$ from the center of the parent dynamic fractal, seen as galactic nucleus. Then dynamic fractal dimension is $d(r) = 2$ in the oscillating dynamic fractal. The velocity of basic matter that drives around the stars is proportional to their velocity $v(r) = const$ and hence $M \sim r_i$ because in this case $C = Gm_sM$. This means increase of the inward 3D-spiral vortex of basic matter far from its core, resulting in the observed flat rotational curves of the stars that float in it. The increase is due to accumulation of dynamic fractal elements from the parent galactic type dynamic fractal structure.

There is enigmatic constant force that pulls sunward the Pioneer 10 and 11 spacecraft, observed from 20 to 70 AU [21]. Then from the Equation of unifying interaction (7) we obtain $v^2(r)/r = a_p = const$ for $r$ belonging to the given range, where $a_p$ is the anomalous acceleration, created from 3D-spiral vortex of basic having velocity $v(r) \sim r^{1/2}$. The increase of the velocity of basic matter outward means expansion of the 3D-spirally-faster-inward-oscillating vortex of basic matter, i.e. dynamic fractal, whose core is seen as the Sun and whose outer structure drives the planets around. This expansion is similar to that of the magnetosphere during southward interplanetary magnetic field (IMF), corresponding to the magnetic attraction, shown in Figs. 6 and 7. Then the basic matter and hence the mass $M$ of the expanding vortex increases as $M \sim r^2$ from Equation (10), written for this case $r_i = r$. This is due to direct entry of Galactic basic matter flux $\sim r^2$ into the 3D-spiral vortex of the Sun. Similarly, the magnetospheric expansion (contraction) during southward (northward) IMF [16, 17] may create slight changes in the gravity field vector at low latitudes as the inner faster (outer slower) moving 3D-spiral vortexes of basic matter shift outward (inward) together with the geomagnetic and gravity fields they are likely to create. The prevailing inward (initial Figs. 1A and 3) 3D-spiral vortex of basic matter is indicated as dark matter, accumulated in the oscillating structure and so it increases with aging. This flattens the rotational curves of spiral galaxies. The expansion of the 3D-spirally-faster-inward-oscillating pattern of unifying interaction, due to attraction with a much larger one from the higher hierarchy (parent) structure, creates anomalous constant inward acceleration at atomic scales. This may attract comets and asteroids and account for their highly eccentric orbits. At planetary scales the expansion of dynamic fractal of the Sun, i.e. the increase of the velocity of basic matter outward is likely to push the planets outward. The expansion of the dynamic fractal of the Sun, indicated by the Pioneer anomaly, will push slightly the planets outward. This is likely to create discrepancies between the predicted and observed orbits of the outer planets, currently associated with attraction from Planet X, which is believed to be beyond the orbit of Pluto. The dynamic fractals attract and expand or repel contracting, similarly to the magnetosphere as it couples with the IMF (Fig. 1 in [16] and Figs. 1-3 in [17]) and similarly to the magnetic attraction and repulsion shown in Figs. 6 and 7. The self-similar evolution of the pattern of one seen as fundamental magnetic interaction draws dynamic fractal firework universe having 3D-spiral code (Figs. 6, 7, 1-5).

The ring shape of the Cartwheel galaxy [22] can originate from expansion of its galactic dynamic fractal resulting from attraction with a much larger one. The stars of the Cartwheel galaxy have followed the expanding lines of unifying force similarly to ions from plasma sphere that drift outward during a period of southward IMF created magnetospheric expansion like that shown in Fig. 6. This period corresponds to magnetic attraction occurring between attracting patterns of unifying interaction of Fig. 1B type, observed as the Earth's magnetosphere and the IMF [16, 17]. The stars of the Cartwheel Galaxy, which have been not strongly coupled with its nucleus, moved outward as its dynamic fractal expanded. The bright blue ring of stars far around the galactic nucleus shows that their stellar size dynamic fractals are disturbed in agreement with perturbation created from their outward motion. That is why they will release more intensely agitated much smaller ones, seen as atomic matter, that releases even too much smaller ones observed as light. The 3D disturbance of the velocity field of the Cartwheel galaxy around its dominant circular velocity component of about 217 km/s [22] is likely to be produced in the described 3D-spiral discrete dynamic fractal attraction happening at galactic scales. The creation of the Cartwheel galaxy is likely to require some threshold of direct external flux entry like shown in Fig. 6. This is likely to occur rarely. So Cartwheel is not a typical galaxy.

The size of a dynamic fractal is defined by the velocity of its basic matter $v(r)$ at distance $r$ from its center (Fig. 1). The size of the fractal is proportional to its resistance to acceleration, known as mass $m \sim v^2(r)r$ [2]. The size of the dynamic fractal, whose core is seen as the Sun, is defined by the orbital velocity $v(r)$ of one of its planets at distance $r$ from the Sun. For example, $v(r) = 29.79$ km/s and $r = 149.60 \times 10^6$ km = 8.3 light minutes = $1.6 \times 10^{-5}$ light years for the planet Earth. Similarly, the size of the Galactic 3D-spiral vortex of basic matter (Fig. 1B) is defined by the velocity $v(r)$ of a star from the disk of the Galaxy, e.g. the Sun, at distance $r$ from the Galactic center. The Sun moves around the center of the Galaxy with velocity $v(r) = 250$ km/s at distance $r = 26,000$ light years. Let us roughly assume that the velocities of stars are $v(r) = 250$ km/s $= const$ for $50,000 \geq r \geq 26,000$ light years from the center of the Milky Way Galaxy. Then the mass of the basic matter found in the Galactic dynamic fractal is equal to that of about $[250^2 \times 5 \times 10^4]/[29.79^2 \times 1.6 \times 10^{-5}] = 2.2 \times 10^{11}$ times the mass of a secondary dynamic fractal that has the size of the Sun. Therefore the vortical motion of the Galactic dynamic fractal can drive a similar amount of basic matter in form of secondary dynamic fractals, observed as about $2 \times 10^{11}$ sun like stars that build the visible matter of the Milky Way Galaxy (Fig. 1B, Fig. 3). This rough quantitative assessment shows that essential part of the Galactic fractal has already unfolded. Hence the Milky Way Galaxy is not a young. It has allowed time for the small stars to cool and to become planets. Time was also allowed for the evolution of life. *The logic of the dynamic fractals unfolding is irrefutable like a proven theorem. Whatever exists attracts itself and bounces back (Figs. 1-3). This happens at self-defined scales and so creates the finite sources of all-building unifying interaction (Figs. 4, 5). The fundamental elements of the dynamic fractal firework universe are ejected from the oscillating denser insides of similar finite larger ones. Afterwards they move around driven by the 3D-spiral structure of their sources as shown in Fig. 3. That is why space bodies moving around larger ones are seen, e.g. stars move around galactic nuclei and atoms move around the centers of stars and planets. This accounts for the observed rotation of the latter around their axes.*

Every change (fragment) of the dynamic fractal is also a dynamic fractal, fundamental element, which depending on its size is seen as time, space, body and light. Time is equivalent to the change of the volume (size) of the dynamic fractal (Fig. 1). Hence time is a dynamic fractal that is a local change in the fractal structure. Time is local originating from each 3D-spirally-faster-inward oscillating dynamic fractal element of reality. Nature is made of similar multi-scale oscillators whose frequencies increase inward and they self-organize (synchronize) to unfold [23]. The change and lack of change of the dynamic fractal at the scales of observation are seen as time and space, respectively. Space and time is how we see the fundamental dynamic fractal elements, i.e. the 3D-spiral vortexes of basic matter called protobodies [2]. Therefore space and time are not fundamental (not initial) enough for the description of the universe, more fundamental is what creates what we see as space, time and everything else [2]. Space and time are perceived as continuous because the elements creating the perception of space and time are integrated, like movie in ones brain [2].

The universal scale created by the dynamic fractal (Fig. 1) allows obtaining of universal equivalence, i.e. proportionality, between two objects expressed in its terms. Einstein's energy and mass equivalence is simply obtained from this universal equivalence [2]. The dynamic fractals are ordered along their hierarchy of origin. The finite larger dynamic fractals create similar smaller ones (Fig. 3). The result is a complete and self-consistent picture of one existing and unique dynamic fractal firework universe. It evolves self-similarly though cycles, which annihilate everything else because the most initial 3D-spiral vortex collapses (Fig. 1A) on its way to begin to oscillate faster inward and thus to eject similar ones as shown in Fig. 1B and Fig. 3. The latter do the same and so on. Thus the fundamental elements of the firework universe are created (Figs. 1-5). The firework universe evolves self-similarly and so it is always governed by similar laws of physics, which are expressed in the equation of unifying interaction (Eq. (7)).

Size is everything in the dynamic fractal structure. The size of the dynamic fractal $v^2(r_i) \times r_i$ is nearly constant $C$ in the oscillating structure at the scales of its existence. This constant is characteristic for the considered dynamic fractal and has to be proportional to its resistance to acceleration $m$, called mass. The Equation of unifying interaction (7) for oscillating structure having parameters $d(r_i)=2$ and $C_i = GmM$ leads to

$$v^2(r_i) \times r_i = C_i \sim M, \qquad (10)$$

where $v(r_i)$ is the velocity of basic matter that builds the considered dynamic fractal (Figs. 1B, 3) at distance $r_i$ from its core, $M$ is its resistance to acceleration (mass). The smaller bodies, i.e. dynamic fractals, follow the velocity of the basic matter that builds the larger one (Fig. 3). Then $v(r_i) = 2\pi \times r_i/T(r_i)$, where $T(r_i)$ is the time it takes for the basic matter to rotate at angle $2\pi$ in its 3D-spiral fractal structure. So it drives around its core a smaller body (e.g. a planet) with a period $T(r_i)$, similarly to the secondary one driven around its source as shown Fig. 3. It can be seen from Fig. 1B that the velocity of the basic matter $v(r_i)$ at distance $r_i$ from the 3D-spiral dynamic fractal core draws nearly circular trajectory. Then Equation (10) written for any two planets turns into the third Kepler's law – $T^2(r_i)/r_i^3 = const$, where $T(r_i)$ and $r_i$ are the sidereal periods of the planets and their mean distances from the Sun.

Let us consider Equation (10) for the dynamic fractals of the Sun and the Earth and calculate the ratio $R_{SE}$ between their masses. The bodies are driven around with velocities, which are proportional to the velocity of the basic matter that builds the dynamic fractal in which they drift as shown in Fig. 3. Hence

$$R_{SE} = v^2_S(r_{iS}) \times r_{iS} / v^2_E(r_{iE}) \times r_{iE} = 3.30 \times 10^5,$$

where $v_S(r_{iS}) = 30$ km/s is the velocity of the Earth driven around by the dynamic fractal of the Sun at distance $r_{iS}=1.459\times10^8$ km from its core seen as star (Fig. 3). Similarly, for the dynamic fractal, whose core is seen as the Earth - the first space velocity is $v_E(r_{iE}) = 7.90$ km/s at $r_{iE} = 6.378$ km from the Earth's center. The calculated ratio in terms of dynamic fractals (protobodies) between the masses of the Sun and the Earth is $3.30\times10^5$. This is "strikingly equal to the reciprocal mass of the Earth $3.29 \times 10^5$ [2]". The equality to the reciprocal mass of the Earth will hold up to all digits of accuracy, after using more precise orbital data, because the application of the equation of unifying interaction at the considered scales converges into Newton's law of gravity. *Therefore the Sun and the Earth are stellar type dynamic fractals or in other words stellar type protobodies. The Earth is a small cooled star, made of huge stellar type nucleus, which is the source of its thick atomic shell we live on.*

Table 1 shows the orbital data for solar system planets. It confirms their motion in the dynamic fractal of the Sun according to the equation of unifying interaction (Eq. (7)) applied at the considered scales. Here $C_P$ is the calculated $C$ in the Equation (10) and $M_i$ is the mass of the Sun obtained from the orbital data $r_i$ and $T(r_i)$ for given planet.

**TABLE 1.** The orbital data for solar system planets considered in dynamic fractal terms.

|  | Mean distance from the Sun $r_i$ (AU) | Sidereal period $T(r_i)$ (years) | $C_P = [2\pi \times r_i/T(r_i)]^2 \times r_i$ $C_P \sim M_i$ | $C_P/C_{Venus}$ | Orbital eccentricity | Inclination to the ecliptic (°) |
| --- | --- | --- | --- | --- | --- | --- |
| Mercury | 0.3871 | 0.24085 | 39.4762 | 1.0001 | 0.2056 | 7.004 |
| Venus | 0.7233 | 0.61521 | 39.4701 | 1.0000 | 0.0068 | 3.394 |
| Earth | 1.0000 | 1.00004 | 39.4768 | 1.0002 | 0.0167 | 0.000 |
| Mars | 1.5237 | 1.88089 | 39.4754 | 1.0001 | 0.0934 | 1.850 |
| Jupiter | 5.2028 | 11.8623 | 39.5124 | 1.0011 | 0.0483 | 1.308 |
| Saturn | 9.5388 | 29.458 | 39.4852 | 1.0004 | 0.0560 | 2.488 |
| Uranus | 19.1914 | 84.01 | 39.5383 | 1.0017 | 0.0461 | 0.774 |
| Neptune | 30.0611 | 164.79 | 39.4923 | 1.0006 | 0.0097 | 1.774 |
| Pluto | 39.5294 | 248.54 | 39.4755 | 1.0001 | 0.2482 | 17.148 |

It is seen from Table 1 that the Sun behaves like a dynamic fractal, i.e. 3D-spiral vortex of basic matter (Fig. 1B), that obeys the third Kepler's law. The mass of the Sun evaluated from the orbital data of the planets varies from 0.01% to 0.17%. The planets Venus, Earth and Neptune have nearly circular orbits as it is seen from their small orbital eccentricities. The slightly stronger inward 3D-spiral vortex of basic matter that builds the dynamic fractal of the Sun is likely to drive most of the planets. They all move prograde. Then Sun's inward (initial and hence stronger) vortex swirls 3D-spirally prograde. Therefore the mass of the Sun inferred the three planets that have nearly circular orbits will be smaller for the inner Venus and largest for the most outward planet. This dynamic fractal inference is confirmed from Table 1. It results from the relative increase of the inward 3D-spiral vortex far from the core of the dynamic fractal of the Sun due to accumulation of fractal elements (basic matter) from the initial Galactic structure. The accumulated dynamic fractal elements, i.e. basic matter, from the 3D-spirally contracting and expanding initial structure, are transformed into smaller ones some are seen as atomic matter built Sun. The orbital data (Table 1) suggests that the Sun is 0.06% and 0.04% more massive as calculated, respectively, from the nearly circular orbits of Neptune and Venus and Neptune and Earth. This is due to outward relative increase of the initial (inward) 3D-spiral vortex of basic matter that builds the dynamic fractal, i.e. the vortex of basic matter (Figs. 1-3), in whose core we find the Sun.

The dynamic fractals account for the nature dark matter far from galactic nuclei, stars and planets. A dedicated spacecraft mission is necessary to explore this prediction around the Earth and for the other bodies in the solar system. A set of prograde and retrograde satellites, having equatorial circular orbits at far different altitudes (e.g., about $2 \times 10^3$ km and $3 \times 10^5$ km) will show existence of dark matter near the Earth after processing their orbital data in way similar that shown in Table 1. This will confirm the existence of the Earth's stellar type dynamic fractal (Fig. 1B), made of 3D-spirally-faster-inward oscillating basic matter.

Every body is ejected from and moves around its source (Figs. 1-5). The much smaller ones, seen as light, leave the discrete 3D-spirally-faster-inward-oscillating structure of their source. Then observer in the cosmic cloud, shown in Fig. 5, will find similar bodies (dynamic fractals) everywhere. That is why the near and most distant parts of the accessible universe look alike. The surprising discoveries of galaxies, clusters of galaxies and heavy elements in the far distant universe [e.g. 4-6] are in agreement with dynamic fractal firework universe having 3D-spiral code (Figs. 1-5). The fundamental dynamic fractal elements are made of one basic matter and create unifying interaction (Eq. (7)). They simply account for the puzzling nature of dark matter. Only the dense insides of the atomic size dynamic fractals are visible at the tiny scales of the much smaller ones, seen as light ejected from the discrete 3D-spirally-faster-inward-oscillating structure of the atoms.

Arp suggested that quasars are ejected from galactic nuclei [24, 25]. Figure 3 describes a mechanism for this ejection, when it is considered at the galactic nucleus (initial) and the quasar's (secondary) dynamic fractal scales.

The initial nucleus becomes perturbed and ejects secondary ones, most likely in the moment of its creation as it starts to oscillate and comes from its source (Fig. 3). This universal mechanism depending on its scales can eject quasars and/or galaxies. Some quasars are likely to evolve into galaxies as their oscillating dynamic fractals accumulate basic matter from their parent fractal structure and eject it to smaller scales, seen as stars. The Earth's magnetosphere contracts and expands [16, 17]. Thus it accumulates energy from its environment and frees it at smaller scales, seen as magnetic disturbances and auroral displays [16, 17]. So does everybody at scales of its own (Fig. 3). There are many observations of high redshift quasars and/or galaxies, which are nearby lower redshift parent galaxies [e.g. 26-29]. The higher redshift of these secondary objects suggests that their light size dynamic fractals are less dense. Thus they expand faster than the dynamic fractals of light that come directly from atomic size ones, created from stellar type dynamic fractals, born directly in the nucleus of the parent galaxy. The stars created from the nucleus of the parent galaxy are likely to emit slower redward shifting light than the stars, cast from nuclei ejected from the parent one. Everything increases source ward. So it becomes denser source ward (Fig. 3) and hence it is likely to expand slower. This elucidates the puzzling nature of the intrinsic redshifts. The redshift discrepancy can be quantified and used as indication of origin.

The dynamic fractal elements or simply the dynamic fractals, seen as galaxies and quasars, are ejected from the cores of bigger galactic size dynamic fractals and from the cores of the sources of the huge galactic fractals. Then there may be some residual perturbation in the velocity field of stars of parent galaxies and/or activity of their nuclei. It is interesting to look for signatures in the velocity field associated with active galactic nuclei. The distribution of quasars having given redshift, which are not associated with parent galaxies, will correspond to the distribution of the sources of the galactic size dynamic fractals. These quasars are likely to be recently ejected "baby" galaxies. That is why quasars are found far from the old Milky Way Galaxy [2].

The 3D-spirally swirling basic matter builds light (Figs. 1B, 3) and leaves the dynamic fractals of the quasars. So the dynamic fractals of light travel through relatively denser basic matter regions, accumulate more basic matter, expand faster and thus create higher redshift. The latter corresponds to interaction described with a longer wavelength of light. The dynamic fractals (protobodies) show wave-particle properties [2]. The discrete interaction between the light size dynamic fractals and the atomic size ones, which build a tiny slit, creates the observed bright and dark bands on a screen behind the slit. The dynamic fractals of light pass through the center of the slit and create the central bright stripe. The others are discretely bent by the inward 3D-spiral vortexes of basic matter that builds the slit or decelerated by the outward ones and then accumulated into the dynamic fractal of the slit, which is topologically equivalent to that shown in Fig. 1B. Hence it preserves its properties by differing only in size made by adding of similar dynamic fractal elements. The bending of the trajectories some photons and the deceleration and accumulation of others produce bright and dark lines the screen behind the slit. In this way single light photons (i.e. light size dynamic fractals) passing through two slits during a long period of time and a short light beam will create the observed same pattern of bright and dark lines on the developed photo sensitive material of the screen. There is no need of strange self-interference. Thus one of the great mysteries of the micro world is solved. A similar process working at the scales of the stellar type dynamic fractals of the planets is likely to create the gaps in the planetary rings. Some particles of the rings are driven along their trajectories by the inward 3D-spiral vortex of basic matter that builds the planets (Fig. 1B the inward pointing arrows) others are decelerated in the weaker outward (secondary) vortex and so removed from trajectories seen as gaps.

The light size dynamic fractals originate from the 3D-spirally-faster-inward oscillating insides of the atomic ones as shown in Fig. 3. The light size dynamic fractals are small enough and so leave the vibrating discrete (open) 3D-spiral structure of their source rather than being driven around its core. The tiny dynamic fractals, i.e. the tiny 3D-spiral vortexes of basic matter, seen as light, originate with discrete velocity $v(r_l)$, whose dispersion is confined into the hyper micro scale $r_l \ll r_a$, where $r_l$ is the scale of light created in the point of its ejection from the oscillating atomic size dynamic fractal and $r_a$ is the scale of the latter that creates to the accuracy of observation. Equation (10) written for the dynamic fractal of light gives

$$v^2(r_l) \times r_l = C \sim m = 0. \tag{11}$$

In the fundamental framework of the dynamic fractals everything is finite (discrete) due to its 3D-spirally contracting and expanding structure. There are no zero, no infinities and no singularities in the dynamic fractal fundamental framework. There are numbers like $C \neq 0$ that correspond to finite (discrete) dynamic fractal elements. The observed zero mass of light photon suggests that $r_l$ is too small but finite, i.e. not infinitely small. Hence $r_l$ can converge into a constant in a set made of finite, i.e. not arbitrary large or small elements (numbers). Then from Equation (11) follows that $v(r_l) = const$. More precisely $v^2(r_l)$ has dispersion that corresponds to the dispersion of $r_l$, which is below the accuracy of observation performed at the much larger atomic scales. *Therefore the zero rest mass* *requires observation of constant speed of light size dynamic fractals in vacuum. Their expansion on the long way from the distant galaxies creates the intrinsic nature of the cosmic red shift.*

The light photons leave the atoms with discrete velocities whose dispersion is too small to be detected at atomic scales. Afterwards light travels the space that is built from the atoms of the Michelson-Morley interferometer. This space also remains unchanged during the light's travel. This space remains the same after imaginary removal of the rest of reality. Hence a constant velocity of light in all directions will be observed. Similarly, a constant velocity of sound is found a flying iron bar [2]. The perturbation and the medium of its propagation are confined within the interferometer or the bar. Therefore the velocity of the perturbation, light or sound, is independent of the motion of the interferometer or the iron bar [2].

The size of the dynamic fractal is created in a hierarchy of origin shown in Figs. 1-5. It is quantified with universal scale according to Equation (10) considered in mass terms. The basic matter moves faster at distance $r$ from the center in the larger, i.e. more massive, dynamic fractal. The decrease of universal scale, i.e. the decrease of mass in Equations (10), leads to decrease of everything including its dispersion. This leads to observation of a constant speed of light in vacuum at the accuracy provided by the much larger atomic size dynamic fractals. The dispersion increases with scale. For example, the ratios between the masses of the lightest and heaviest chemical elements, smallest and biggest stars and little and largest galaxies are respectively $\sim 10^2$, $\sim 10^7$ and $\sim 10^8$, assuming that planets are small cooled stars and galaxy is every assembly of stars (e.g. globular cluster) moving around a very massive object. The latter is the core of galactic type dynamic fractal that has ejected stellar ones. So its mass is assumed to be proportional to the number of stars that orbit it. The dispersions of the masses of the stars and the dispersions of the masses of the chemical elements in the globular clusters have to be smaller than those in galactic disks because of the smaller universal scale (size) of the globular clusters allows smaller dispersions. That is why low abundances of heavy elements are observed in the globular clusters. The discovery of intermediate size black holes, i.e. very massive objects, in the globular clusters [30-32] is experimental confirmation of their dynamic fractal origin.

There are plenty of evidences that the Earth expands [e.g. 33-37]. The failure of expansion theories is due to inability to suggest a mechanism providing the energy of expansion. Hugh Owen argues that the inner core of the Earth "is in some state nobody has yet imagined, a state that is undergoing a transition from a high-density state to a lower density state, and pushing out the crust, the skin of the Earth, as it expands [38]." That is just what the creation of atomic size dynamic fractals in cores of the initial stellar type ones does (Fig. 3). The basic matter evolves self-similarly from higher to lower density states. This accounts for the creation of intrinsic cosmic redshift, expansion of the Earth and ejection of stars and quasars from galactic nuclei (Figs. 1-5). One may take a globe of the Earth in his hand and consider the fitting of the coast lines of South America and Africa by decreasing the radius of the globe and some rotation of these continents. Radial motion plus rotation draw a spiral curve. This spiral can be created from the 3D-spirally-faster-inward-oscillating basic matter of the stellar type dynamic fractal of the Earth, while it drives around and outward its growing thicker atomic shell. The 3D-spirally-faster-inward-oscillating basic matter that builds the dynamic fractal of the Earth is much denser in its inner core than the atomic matter. The super dense core of the Earth's dynamic fractal 3D-spirally oscillates and ejects less dense atomic nuclei. Thus it creates the mighty energy source for the Earth's expansion.

In the 3D-spirally-faster-inward-oscillating dynamic fractals the initial inward 3D-spiral swirl of basic matter prevails and creates universal attraction, while the 3D-spiral outward swirl creates universal repulsion. The inward swirl of basic matter that builds the dynamic fractal of the Earth attracts bodies to its surface and so creates Earth's gravity. Newton's law of gravity works well in the ambient space simply because it is derived from the equation of unifying interaction drawn for oscillating structure (Eq. (7) for $d(r_i) = 2$) and properly obtained constants at these scales, corresponding to the masses of the Earth, the Sun and solar system planets. The masses are proportional to the ability of the dynamic fractals to drive around smaller ones (Eq. (10)) with acceleration proportional to the inverse square of the distance from their cores (Eqs. (10) and (7)). The circular orbits of the moonlets around the rotating huge ellipsoid 87 Sylvia asteroid [39] are great puzzle when considered in the terms of Newton's law of gravity. If this law fails in a triple asteroid system, then what can be said for the rest of the universe?

The bulk of the Sun is made mostly of Fe, Ni, O, Si and S, suggesting "that fusion of hydrogen is probably not the Sun's primary energy source" and origin of the solar system from radioactive supernova debris [10, 40, 41]. The bodies in the firework universe have explosive beginnings and ends as they are ejected from their sources and later are annihilated by them (Figs. 1-3). The ejection of atomic matter from the cores of the stellar type dynamic fractals of the sun and the planets, created the highly radioactive beginning of the solar system. The creation of new atomic matter in the centers of stars and planets heats their insides and increases their radii. Thus Earth expands and the huge Pangea landmass was torn to what became modern continents. The Earth's nucleus creates atomic nuclei and thus powers the lava upwelling mid ocean ridge and volcanic activity. The unfolding nucleus drives slower its growing a thicker atomic shell (Fig. 3). Hence the Earth's rotation around its axis will slow down. This prediction is in agreement with findings of increase of the length of day as the Earth ages [e.g. 42]. Similarly, on much smaller scale the expansion of the structure of the aging light that arrives from the distant galaxies creates the cosmic redshift [2].

The solar wind dynamic pressure and the interplanetary magnetic field make the magnetosphere to 3D-spirally contract and expand, thus indicating the 3D-spiral structure of the seen as fundamental magnetic interaction [16, 2, 17]. The stellar type dynamic fractal, i.e. 3D-spiral vortex of basic matter, seen as the planet Earth, creates its magnetosphere and produces new atoms from the nucleus of our planet. Then the global expansions and contractions of the magnetosphere during magnetic storms are likely to intensify the creation of new atomic nuclei in the center of the Earth. This can increase of the frequency of earthquakes.

The all-building 3D-spirally-faster-inward-oscillating dynamic fractals (Fig. 1B) attract each other and keep their cores apart. This creates universal attraction and repulsion across the scales of nature and so accounts for the mysterious cosmic repulsion. The latter prevents dense populations of stars from merging into one huge mass. Stars couple like atoms on stellar dynamic fractal scales. The galactic dynamic fractals interact similarly at their scales and create the puzzling stability of galactic clusters. The universe is dynamic fractal having 3D-spiral code (Figs. 1-5). Its dynamic fractal dimension changes from 0 to 3 (Eq. (7)). Thus it accounts for Olbers' paradox and the fractal dimensions of observed galaxies distributions [8]. The obtained fractal dimensions about 1 - 2 of galaxies distributions [43, 44 and the references therein] confirm the dynamic fractal firework universe, indicating existence of parent galactic dynamic fractals associated with Great Attractor [15] type objects. If the galaxies were tightly coupling, not separated by the space created from the more initial dynamic fractal elements, as it is shown by the flat velocity distributions of stars far from the galactic nuclei, then the fractal dimension of galaxies distribution will be close to 3.

The asteroid 87 Sylvia is discovered to have two small moonlets moving around at 710 ± 5 and 1,360 ± 5 km with periods 1.3788 ± 0.0007 and 3.6496 ± 0.0007 days on equatorial, circular and prograde orbits [39]. The mass density of the asteroid, estimated from the orbital data is 1.2 ± 0.1 g/cm$^3$, suggesting "a rubble-pile structure with a porosity of 25–60 per cent" [39]. The shape of 87 Sylvia primary is an ellipsoid with axes $a$ = 192 km, $b$ = 132 km and $c$ = 116 km and a mean radius $R_e$ =143 km [39]. It is very puzzling how the nearly circular orbits of the moonlets originate in the gravity field close to huge ellipsoidal porous, ruble-pile, central body, which rotates faster around itself than the time it takes for the inner moonlet to complete one orbit. In this way 87 Sylvia tends to be torn by centrifugal force. It can be written $v(r_i) = (2\pi \times r_i)/T(r_i)$, where $v(r_i)$ and $T(r_i)$ are the velocity and the period, i.e. the time for rotation at angle $2\pi$ = 6.2832, of the basic matter at distance $r_i$ from the center of the considered 3D-spiral vortex (Fig. 1B) in which 87 Sylvia and its moonlets are likely to rotate. Then from Equation (10) and the orbital data $r_1$ = 710 km, $T(r_1)$ = 1.3788 days, $r_2$ = 1,360 km, $T(r_2)$ = 3.6496 days we calculate $v(r_1)=(2\pi \times r_1)/T(r_1)$, $v^2(r_1) \times r_1 = C_1$, $v(r_2)=(2\pi \times r_2)/T(r_2)$, $v^2(r_2) \times r_2 = C_2$ and $C_2/C_1$ = 1.0031. Therefore Equation (10) holds at level $C/C = m/m = C_2/C_1$ = 1.00. This indicates the existence of the considered dynamic fractal, whose 3D-spiral vortex of basic matter drives the observed motions 87 Sylvia and its moonlets. The obtained $C_2/C_1$ = 1.0031 suggest 0.31 % more mass (dark matter) accounts for the motion of the outer moonlet than for the motion of inner one. This is a slight indication, deserving further investigation and probably not distorted by the accuracy of $r_1$ and $r_2$ measurement, that the considered interaction resembles the one that flattens the rotational curves far from galactic nuclei, due to accretion of dynamic fractal elements from the similar initial (parent) structure. The holding of Equation (10) at level 1.00 clearly indicates that moonlets are driven in one 3D-spirally-faster-inward-oscillating basic matter swirl. Table 1 shows that the motion of the moonlets is similar to that of solar system planets. The dynamic fractal structure is confirmed in both cases with accuracy 1.00. The structure rotates faster inward (Fig. 1B). This drives 87 Sylvia faster around itself than the periods of its moonlets, found further from this huge asteroid, which is in the center of the swirl. The observed in nearly one plane circular orbits of the moonlets [39] are consistent with their motion in a 3D-spiral vortex of basic matter, called dynamic fractal or protobody (Figs. 1-5).

The asteroid belt originated from an exploding planet as its atomic matter shell collapsed and became annihilated from its source [2]. The nucleus of the planet created a thick atomic shell around itself. Afterwards the nucleus contracted at atomic scales and annihilated a part of its atomic cover in a transition like that shown in Fig. 1B → 1C [2]. The very dense and massive, collapsed stellar type, nucleus of the planet fell on the Sun and annihilated itself. In this way perturbations, i.e. smaller scales dynamic fractals, were created in the dynamic fractal of the Sun. The perturbations traveled deep into the initial fractal and afterwards outward as shown in Fig. 3. The created secondary 3D-spiral dynamic fractals, i.e. vortexes of basic matter (Figs. 3 and 1B) drive the rotational motions the asteroids, produced from the supernova type explosion of the planet. So binary and multiple asteroid systems were created. There has to be a spacecraft mission to 87 Sylvia to check its density by measuring the velocity of sound waves in this asteroid. The prediction is that the real density of the asteroid is much larger than the calculated $1.2 \pm 0.1$ g/cm$^3$ in the standard framework. It is very interesting to look for unifying force disturbances, i.e. 3D-spiral vortexes of basic matter. This "basic turbulence" will appear at these scales as small gravity perturbations that create acceleration $\leq v^2(r_2)/r_2 = 5.4\times10^{-4}$ m/s$^2$, which holds the outer moonlet of 87 Sylvia. These perturbations can be observed in a spacecraft that moves across the solar system and particularly into the asteroid belt. There has to be unifying force disturbances having space scales $< 10^3$ km and showing as gravity variations $< 5.4\times10^{-4}$ m/s$^2$ in the asteroid belt. There may be belts of slight "gravity turbulence" around stars and planets, created by secondary dynamic fractals, appearing in the 3D-spirally oscillating initial stellar type ones.

The double asteroid 90 Antiope [45, 46] is made of two bodies each about 85 km across, separated by about 160 km orbiting a point in the midway between them. The asteroid is believed to have been once a single body and is considered to be very puzzling. This double asteroid may have originated like 87 Sylvia. Two bodies were captured in this case by a similar dynamic fractal. It set them in motion, further from its faster moving core. Then their period of rotation will be larger than that of 87 Sylvia, which is the core of its 3D-spiral vortex of basic matter. The double asteroid Antiope rotates for about 16 hours. Sylvia spins once for about 5 hours. *The puzzling triple asteroid system 87 Sylvia and the double asteroid Antiope move as if they are merged into 3D-spiral vortex of basic matter (Figs. 1B and 3).*

**Some testable predictions:** 1) the gravitational constant G depends on the nature and temperature of the coupling masses from which it is obtained and on the distance between them, because the dynamic fractal dimension in the equation of unifying interaction decreases with distance as the larger space intervals are created from a smaller number of initial (parent) dynamic fractals; 2) everything expands and contracts at scales of its own – the geospace oscillates at the time scales of magnetospheric expansion and contraction – minutes, hours and days because the magnetosphere originates from the stellar type dynamic fractal of the Earth, the ambient space oscillation can be obtained from analysis of the trajectories of the GPS satellites and/or gravimetric studies; 3) the distance between the Moon and the Earth also oscillates and a special project is necessary to find out the frequencies of this intrinsic oscillation and to use it as natural hazard precursor; 5) there should be some correlation between geomagnetic and gravitational disturbances [16] because they are manifestation of one unifying force (Eq. (7)) measured at different finite (discrete) scales; 6) there should be dark matter around the space bodies - planets, stars, clusters of stars and galaxies created from the increase of the inward 3D-spiral swirl of basic matter there, resulting from accumulation of basic matter from the initial (parent) structure; 7) the angular velocity of the inmost core of the Earth is greater than that of its crust because every motion is driven by faster inner (initial) one (Figs. 1 and 3); 8) strong magnetic storms will be followed after about 10 days (from my rough preliminary assessments) by global increase of the number of strong earthquakes (magnitude > 4.5) because the magnetic storm is expansion and contraction of magnetosphere [16, 17], hence the dynamic fractal of the Earth expands and contracts and so ejects more secondary ones (atomic nuclei) from its core; 9) the distribution of quasars having given redshift and not belonging to parent galaxies will be patchy - the angular scale of the patches will decrease with the increase of the distance to these quasars as measured by their redshift increase; 10) highly sensitive gravimeter on a spacecraft flying within the asteroid belt will measure variations $\leq 5.4\times10^{-4}$ m/s$^2$, caused by 3D-spiral vortexes of basic matter, having scales less than $10^3$ km, created in the oscillating similar structure of the stellar type dynamic fractal, in whose kernel we find the Sun; 11) 87 Sylvia asteroid is a solid rock body and so is its density is much larger than 1.2 g/cm$^3$ because the asteroids are chunks of exploded planet; 12) there is universal source ward anisotropy, e.g. anisotropy of galaxy distribution is likely to increase with increase of their redshift; 13) the relative density of craters per unit area into the larger ones will show when the larger formed before or after the smaller ones; 14) the distribution of craters will show if they resulted from meteorite impacts or bubbles on surface of cooling small planet seen as planetary moon; 15) Figures 1-5 show that an observer from a galaxy will find similarity between the near and distant dynamic fractal universe, the observed universe is like a field planted with burning candles (galaxies) and so it will show no expansion at the scales of observation.

There is a dark matter around the Earth. The gravitational constant will show dependence on the distance between the spheres in the experiment for its determination. The outer spacecraft at equatorial circular orbit will move slightly faster than what is expected from the inverse square Newtonian law of gravity. The rescaling of the orbital data for the Earth, Venus and Neptune will provide quantitative assessment of this prediction. The elliptical and spiral galaxies are unlikely to be well mixed. The unifying force Equations (7, 7') can account for the formation of the gaps in the planetary rings, their formation and for the recurrent appearance of spikes in them. The particles of the rings came most likely from exploding planetary like moons [2]. Correlation of the spikes of Saturn's rings with its strong magnetic storms and equatorial wind velocities is likely.

The indication for isotropic distribution of craters will pose a deep problem for the origin of the meteorites that pounded the surface of the moons from all directions. What kind of impact creates the observed upwelling in the center of some moon craters?

There is a dark matter around stars and planets, indicated by the orbital data of near and distant satellites of these space bodies, explained with the equation of unifying interaction, applied at the considered dynamic fractal scales (Eqs. (7) and (10) and Table 1). The double asteroid 90 Antiope consists of two parts having size of about 85 km, separated by 160 km and rotating around a point, which is the middle of seen as empty space between them. The moonlets of 87 Silvia are on nearly circular orbits, which are close to the faster inward rotating ellipsoid primary body. The gravity fields of these asteroids can hardly account for their observed motions. It seems that Newton's law of gravity has run out of the scales of its successful application. The observed motions of these asteroids suggest that they are merged into and driven by a 3D-spiral vortex of basic matter, which originates in the similar much larger one in whose core we see the Sun (Fig. 3). Similar vortexes are likely to be found in the asteroid belt as indicated from expected disturbances in the gravity force in the order of $10^{-4}$ m/s$^2$ at space scales equal and less than $10^3$ km. This gravity turbulence, created by 3D-spiral vortexes of basic matter, is likely to be frequent in the asteroid belt and probably in other belts around the Sun and at smaller scales in belts around the planets. Maybe the appearance of such secondary vortexes between the orbits of Mars and Jupiter has led to faster unfolding of the stellar type dynamic fractal, which would had been seen as a planet there. The consequent supernova like explosion of this planet created the asteroid belt. The two parts of 90 Antiope may have been captured from the debris of such explosion, which emitted very strong radiation as the inner part of the planet was annihilated from its collapsing at atomic scales stellar type nucleus.

The gravity field of the Earth will show small variations that have intrinsic nature and occur at global magnetospheric perturbations time scales of minutes, hours and days because gravity and geomagnetic field originate from measurement at different scales of the stellar type dynamic fractal (protobody) creating the Earth. The evolution of the discrete dynamic fractal structure of the Earth is likely to create the reversals of its magnetic poles during its history.

The sunward force acting on the Pioneer spacecraft [21] is confirmation of testable prediction 6). The faster rotation of stars in the Galactic plane than above and below it is similar to the faster equatorial motions of the gas giant planets and the Sun. The faster rotation in the equatorial plane of the initial dynamic fractal originates from alignment of the secondary ones with the initial dynamic fractal structure and so increase of their interaction with it.

The discovery that the "Earth's inner core is rotating faster than the mantle and crust [47]" confirms prediction 8). Every body moves faster inward because of its dynamic fractal structure does the same (Figs.1-3). It moves 3D-spirally-faster-inward and thus becomes denser and so visible inward. The faster rotation of the Earth's inner core is in agreement with the existence of the stellar type dynamic fractal of our planet and its faster inward rotating nucleus (Fig. 1B). The atomic matter originates from this enormous nucleus and creates the lava upwelling mid ocean ridge and the Earth's expansion [2]. The creation of new atomic matter from the Earth's nucleus will produce upwelling bubbles. This will make the structure of the inner core lumpy.

The discrete 3D-spirally-faster-inward-oscillating structure of dynamic fractals accounts for the observed discrete (quantum) properties of the micro world if the pattern of the unifying interaction is to be scale invariant. The discrete (quantum) properties of the space bodies occur at their time scales, which are much larger than that of observation. So we see continuous motion of stars and planets. It can be predicted that the Earth-Moon and the Earth-Sun distances will show intrinsic variations at the time scales of the orbital periods of these space bodies. Imaginary huge planet entering the solar system with a high speed will change discretely the orbits of planets like an electron shot into an atom.

### 4. DISCUSSION

The buildup of the 3D-spiral dynamic fractal leads to increase of the number of its self-similar elements creating a universal distance *r*. Then the dynamic fractal dimension increases from 0 to 3 with the dynamic fractal density increase as shown in Fig. 1A. Afterwards the dynamic fractal starts to oscillate 3D-spirally-faster-inward. So it remains always finite and thus existing (Fig. 1B). The dynamic fractal dimension drops to 2 in the oscillating structure. This explains the inverse square laws, derived from the equation of the unifying interaction (Eq. (7)). Later the dynamic fractal collapses on its source and turns into a rotating dense remnant (Fig. 1C), which at stellar scales is seen as a rotating dense object, known as neutron star. The multi-scale faster inward contraction and expansion, oscillation, of the basic substance create the cosmic repulsion that prevents the dense populations of stars from merging and keeps the coupling galaxies apart.

One basic matter undergoes dynamic fractal transforms and so accounts for what we see. Space, time, bodies and light are how we see one dynamic fractal firework universe having 3D-spiral code. The firework universe is made of one basic matter that hierarchically unfolds to dynamic fractal sources of unifying interaction. Galactic nuclei, stars, planets, moons, atoms, electrons and light - everything is basic matter 3D-spirally falling into and bouncing from itself. Everything is driven by a faster inward motion (Figs. 1-5). Everything moves faster inward and so it is harder and usually visible inward.

The origin of the universe and its space bodies are described in initial and secondary unifying interactions, swirls of basic matter, which create one dynamic fractal structure, whose self-similar elements remain always finite in the least number of dimensions – 3D that allow this to happen as shown in Figs. 1-5. The resulting firework universe unfolds running downstream because smaller fundamental dynamic fractal elements are ejected like fireworks from the insides of similar finite larger ones (Figs. 1-5). The created multi-scale sources of dynamic fractal unifying interaction build the observed clumpiness of matter at every scale.

The big bang universe runs upstream. It creates everything from seen as elementary particles born from a mysterious matter-antimatter asymmetry that appeared in even more puzzling singularity allowing universe beginning. The big bang picture of the universe is incomplete. It is uncertain what was before the big bang, why it happened at all, and the universe unfolding is controversial. The firework universe is cyclic, singularity free and complete. It is always governed by similar laws of physics described by the equation of unifying interaction, which is drawn in the fundamental dynamic fractal framework (Figs. 1-5 and Eq. (7)). It is shown that the equation of unifying interaction converges into the inverse square laws and the principle of uncertainty at laboratory scales. Hence these well experimentally tested laws and principle are quantitative and qualitative confirmations of the dynamic fractal framework in which the equation of unifying interaction is derived. The equation of unifying interaction is also confirmed with quantitative assessments, based on the orbital data of objects that move around the Earth, the Sun, the nucleus of the Milky Way Galaxies and around the large asteroid 87 Sylvia. The flat rotational curve far from a galactic nucleus is modeled with increase as $M \sim r$ of the outer region of the 3D-spiral swirl of basic matter, in which stars move. This is due to accumulation of basic matter in the galactic 3D-spiral swirl from its parent structure. The obtained about 2 and smaller fractal dimensions of galaxies distributions [43, 44 and the references therein] are also explained in dynamic fractal terms with adding of distance building the fractal elements from the parent structure in which the galaxies move.

The *observed* zero rest mass of light requires observation of its constant speed in way that provides mechanism for intrinsic cosmic redshift. The 3D-spiral swirls of basic matter, called dynamic fractals, create universal scale (Fig. 1). Everything, including its dispersion, decreases toward the smaller scales. This leads to observation of a constant speed of light because the dispersion of its velocity falls bellow the accuracy of measurement performed at the higher hierarchy of atomic scales.

The antimatter properties are produced from 3D-spirally collapsing dynamic fractals of Fig. 1A and 1C type. The 3D-spiral-faster-inward contractions and expansions of basic matter vortexes, called dynamic fractals or protobodies [2], create the observed attractions and repulsions at all scales. This explains the cosmic repulsion that keeps the space bodies from convening in one huge mass.

The circular orbits of the moonlets of the ellipsoidal asteroid 87 Sylvia [39] are consistent with asteroid motions driven by secondary dynamic fractals, i.e. 3D-spiral vortexes of basic matter, appearing in the initial structure, which is the dynamic fractal, whose core is seen as the Sun. The self-similarly evolving dynamic fractal firework universe undergoes all annihilating cycles and so confirms its existence and uniqueness, i.e. there are no other universes, parallel universes, whatever. The firework universe is based on the irrefutable logic of origin (creation or interaction). It shows how everything comes from the insides of similar finite large one. The universe as a whole and every of its bodies are made of finite sources of dynamic fractal unifying interaction. It 3D-spirally-faster-inward contracts and expands and so ejects like fireworks similar finite ones that do the same. The created finite, multi-scale, self-defined and self-similar sources of unifying interaction build the observed lumpy structure at cosmic and atomic scales. The universal self-definiteness of the self-similar contractions and expansions leads to appearance of consciousness that tries to define its source, i.e. to know oneself and the rest of the universe. The 3D-spirally-faster-inward-oscillating dynamic fractal (Fig. 1) suggests that nature is made of faster-inward oscillating elements that synchronize to unfold the interaction (the energy) packed in them. In this way the enigmatic driving force behind self-organization and life is created.

Every problem can be reduced to a deep problem of origin, i.e. what comes first explains everything. The dynamic fractals of unifying interaction, called protobodies [2], create new fundamental framework of revealed origin, i.e. creation of initial and secondary fundamental elements (Figs. 1-5), which account for poorly understood observations from the ambient and most distant space [16, 2, 17]. The mysterious ring of blue stars around the nucleus of M31galaxy [48] indicates their ejection from this nucleus in a process similar to that shown in Fig. 3, working at the galactic scales. The galactic 3D-spiral vortex of basic matter contracts and expands. Thus it recurrently ejects stellar type ones from its nucleus (Fig. 3). They move outward and so create the observed elliptical evolving toward spiral shape of the galaxies. The ring of red stars outward from the blue ones [48] is created as the stars move outward from their 3D-spirally-faster-inward-oscillating source, seen as galactic nucleus. Thus their 3D-spiral vortexes of basic matter (Fig. 1B) become less perturbed due to moving in less dense and slower oscillating parent structure. Then smaller quantities of atomic nuclei are created in their cores and so less heat is released from the cores of the outer less perturbed and so seen as red stars. The latter appear red. The more perturbed are blue. A star can be perturbed by a dynamic fractal element from its parent structure in a way similar the Earth's magnetosphere is perturbed by the southward interplanetary magnetic field. This will lead to contraction and expansion of stellar type 3D-spiral vortex as shown in Fig. 1 in [16, 17] and consequent intensification of secondary, i.e. atomic ones, ejection from its core resulting in stronger heating of its atomic shell. The disks of the blue and red stars "rotate in the same sense and are almost coplanar" [48] indicating motion in a galactic size 3D-spiral vortex of basic matter that follows the rotational sense of the prevailing (initial) 3D-spiral inward vortex. Similarly, the prograde motions of planets are created in a plane close to the ecliptic due to their motion in the stellar type 3D-spiral vortex of basic matter in whose core we find the Sun.

The dynamic fractal firework universe is made of multi-scale similar dynamic fractals. The smaller ones are ejected from finite larger ones and driven around by their outer 3D-spirally contracting and expanding discrete structure. The smallest ones, which are detectable leave their discrete 3D-spirally-faster-inward-oscillating sources and are seen as light, emitted from atomic matter, made of atomic size dynamic fractals. *The structure of reality is discrete, made of finite fundamental dynamic fractal elements, which are integrated like movie frames to create the observed continuity of space and time. In the discrete dynamic fractal structure not all-real numbers have physical meaning because the real numbers map on a continuous space. This fundamentally limits the application of continuous mathematical objects in physics.*

The universe is made of dynamic fractals having 3D-spiral code. The smaller are ejected from the insides of the larger one's and move around in their outer 3D-spirally swirling structure. Every body moves around its faster moving source (Figs. 1-5). Stars move around galactic nuclei, planets orbit stars and atoms move around the faster moving nuclei of stars and planets creating the observed rotation of these space bodies. The small dynamic fractals, seen as light, leave their sources due to the discrete nature of the found all-building unifying interaction. Everything moves faster inward and so it becomes denser inward (Figs. 1-5). The human progress accelerates with deepening of knowledge as faster inward moving dynamic fractal elements are put into practice.

*The parameters in the equation of unifying interaction, i.e. the dynamic fractal dimension $d(r_i)$ and the constants that express the unifying force constant $C_i$ (Eq. (7)), have to be obtained experimentally at different scales. This will require development of many research projects and will quantitatively advance the modeling of nature. The dynamic fractals of unifying interaction provide new fundamental framework for qualitative and quantitative modeling.*

*The fundamental dynamic fractals of unifying interaction run downstream because they eject smaller fractal elements from the oscillating insides of finite larger ones. That is why we cannot build atoms from light photons and similarly we cannot build the universe from seen as fundamental particles without extra assumptions. The firework universe is supported by the principle of parsimony, which requires not doing with more, e.g. with more entities and/or assumptions, what can be done with less. Nothing is less than one thing that appears in different sizes and thus creates the similar elements of the dynamic fractal firework universe.*

Simply speaking, the universe is made of multi-scale nuclei. The smaller nuclei are ejected from finite similar larger ones and move around them. The multi-scale nuclei interact and build the observed clumpiness of matter at every scale. We see only the insides of these nuclei which are dense enough to emit or reflect much smaller ones seen as light. The universe is round, filled with similar round bodies that cast smaller ones from their cores. Every body is made of dynamic fractals that move faster inward. So it becomes denser and usually visible inward (Figs. 1-5). Planets are small cooled stars. The stellar type nuclei (dynamic fractals) eject atomic nuclei and thus account for the hot interiors of stars and planets. The expanding Earth [33-37], the faster rotation of its inner core [47] and the Sun made mostly from heavy elements [10, 40, 41] are in agreement with creation of atoms from the stellar type nuclei of stars, planets and planetary like moons. The discovery of iron rich rigid surface below Sun's fluid photosphere [49] shows that the Sun is a huge hot planet, which is orbited by much smaller and much cooler ones because in their cores new atomic matter is created at much smaller rates that keep only their interiors hot.

The universe is "self-similar on all scales, and, hence, fractal" showing "identity of the observable topology" from $10^{-7}$ cm to up$10^{28}$ cm, made of "coaxially-tubular blocks and blocks of types of cartwheels" [50]. The similarity along the scales of the universe is made directly from similar fundamental elements that build these scales rather than from atomic size objects that mysteriously self-organize in similar patterns having greater and greater sizes. The discovered self-similar skeletal structure of the universe and its tubular and cartwheel blocks [13] are explained with the topological identity of the 3D-spiral swirls of basic matter as they originate from the oscillating insides of their similar sources (Figs. 1-5). The created dynamic fractal firework universe is build from hierarchically ordered along their origin (scale) fundamental elements as shown in Figs. 1-5. It 3D-spirally contracts and expands, thus it remains always finite and opened for interaction with its environment. The interaction of the finite elements shows conservation of some its properties at the scales of observations. The expectations for mechanical properties of the skeletons at cosmic scales – "rigid-body like behaviour of skeleton's constituent blocks" [51] are confirmed with the interaction of the dynamic fractal fundamental elements (protobodies) that show same properties at their scales and so create the functional uniformity of nature [2].

The observed universe is a cosmic cloud of galactic and Great Attractor type nuclei (Fig. 5). They all move around their source similarly to an atmospheric cloud, which moves around the center of the Earth from where most of the atomic nuclei that build the cloud have originated. In this way the observed puzzling similarity between the near and most distant accessible universe is created, i.e. similar space objects are found in the near and most distant universe.

The inverse square laws and their constants are cases appearing at the scales of observation of one unifying dynamic fractal interaction, described with equation, drawn in its 3D-spirally-faster-inward-oscillating basic matter framework. The equation of dynamic fractal unifying interaction simply accounts for the observed flatting of the rotational curves of stars far from galactic nuclei and far from the massive objects in the centers of globular clusters.

## 5. CONCLUSION

The study of the contractions and expansions of 3D-spirally-faster-inward-oscillating vortexes of basic matter, called dynamic fractals, led to the following results:

a) The firework universe is a dynamic fractal whose 3D-spirally-faster-inward-oscillating finite, i.e. singularity free, fundamental elements eject similar secondary ones made of 3D-spiral vortexes of basic matter. The similar elements interact and create similar structures in a revealed hierarchy of origin (Figs. 1-5). The discovered puzzling self-similar skeletal structure of the universe indicating some kind of fractal [13] is so explained.
b) The firework universe runs downhill because it creates smaller fundamental elements in the insides of similar finite ones. The big bang universe runs uphill building everything from seen as fundamental particles, born from a mysterious matter-antimatter asymmetry appearing in a singularity allowing universe beginning. It is very unlikely that small scale elements will self-organize in similar, topologically identical, skeletal structures at atomic and cosmic scales. The big bang universe requires dark matter of uncertain nature to account for the gravitational confinement of the motion of space bodies in cosmic structures.
c) The dynamic fractal fundamental elements of the firework universe are made of one postulated basic matter, whose 3D-spiral contractions and expansions create unifying force described with equation of unifying interaction. The latter turns into inverse square laws - Newton's law of gravity, Coulomb's law, Maxwell equation of electromagnetic field, the principle of uncertainty and Modified Newtonian Dynamics (MOND) like expression at the scales where these laws and principle are found to hold.
d) One basic matter moves 3D-spirally-faster-inward. Thus it attracts itself and bounces 3D-spirally outward. It contracts and expands, oscillates, 3D-spirally-faster-inward. So it accumulates its environment and ejects like fireworks smaller similar ones that do the same. In this way the self-defined, self-similarly evolving, finite fundamental dynamic fractal elements of the firework universe and its 3D-spiral code are created.
e) The motions of the bodies are driven by 3D-spirally-faster-inward moving swirls of basic matter. This is in agreement with the discovered faster rotation of the Earth's inner core, the flat rotational curves of spiral galaxies and is quantitatively confirmed with orbital data for the Milky Way Galaxy, the Sun, the Earth and the triple asteroid system 87 Sylvia.
f) The 3D-spirally-faster-inward moving patterns of basic matter evolve self-similarly through all annihilating cycles and present one self-defined dynamic fractal firework universe, governed by similar laws of physics, expressed with equation of unifying interaction, drawn in the fundamental framework created by the 3D-spiral basic matter swirls, called protobodies or dynamic fractal elements.
g) The third Kepler's law is obtained from the equation of unifying interaction applied to the motions of solar system planets in the 3D-spiral vortex of basic matter in which core we see the Sun.
h) The nature of dark matter is elucidated in terms of bodies driven in the 3D-spiral swirls of basic matter that builds their sources, e.g. galactic size 3D-spirall-faster-inward-oscillating vortexes of basic matter. The prevailing of initial (inward) 3D-spiral swirl appears as dark matter far from its denser and usually visible

       core – galactic nucleus or star. The direct expansion of the 3D-spirally-faster-inward-oscillating dynamic fractal of the Sun due to attraction with similar one from the Galactic structure creates the sunward force acting on Pioneer spacecraft.

i) Dependence of the gravity constant on the nature, temperature and the distance between the coupling spheres in the experiment for its obtaining is predicted. Also existence of dark matter around the stars and planets and association of gravity with magnetospheric expansions and contractions are expected.

    In the dynamic fractal firework universe the fundamental elements of nature are ejected from the oscillating insides of similar finite ones. They all have 3D because 3D is the smallest number of dimension in which the fundamental (initial) elements of the universe remain always finite, ordered along their origin, which is a hierarchy of scale. The created multi-scale dynamic fractal fundamental elements are built from the 3D-spiral motion of one basic matter. The multi-scale elements interact and so account for the observed clumpiness of matter at atomic and cosmic scales. These elements elucidate the nature of dark matter through the equation of unifying interaction, which transforms into Newton's law of gravity at the scales where this law is tested to hold. The logic of the firework universe is the simple logic of one dynamic fractal, i.e. pattern of unifying interaction [2] that self-reproduces itself in different sizes to account for many poorly understood observations and to make testable predictions. The multi-scale dynamic fractal fundamental elements create the observed clumpiness of matter at atomic and cosmic scales and explain the puzzling similarity between the near and most distant universe. The fundamental dynamic fractal elements are likely to cast light also on the subatomic scales with expected radical simplification of the description of the observed interactions there.

    The orbital data for the Milky Way Galaxy, solar system planets, the Earth and the moonlets of the 87 Sylvia asteroid confirm the equation of unifying interaction drawn into the new fundamental dynamic fractal framework. Thus the orbital data confirms the framework and indicates dynamic fractal firework universe made of similar fundamental elements that eject like fireworks smaller ones from their cores (Figs. 1-5).

    The universe is made of similar elements, nuclei, ejected from larger ones in a hierarchy of origin which is a hierarchy of scale. All come from the hyper huge nucleus of the universe. It 3D-spirally contracts and expands and so it remains always finite, i.e. singularity free, in the least number of dimensions – 3D. Thus it ejects smaller similar ones that do the same. The observed puzzling similarity between atomic and cosmic structures [13, 50, 51] is created from similar nuclei having atomic and cosmic scales. The universe is a dynamic fractal having 3D-spiral code. It 3D-spirally-faster-inward contracts and expands, oscillates and so ejects smaller similar elements from the insides of finite larger ones. The universe is made of sources of unifying interaction that eject finite similar ones. The created similar sources of unifying interaction build topologically identical, similar, structures.

    The dynamic fractal firework universe is simple because it is essentially one 3D-spiral pattern of unifying interaction appearing in different sizes. The universe is simple like one thing that self-reproduces itself in different scales. There is fundamental dynamic fractal simplicity. Complexity is created in the process of observation made of too many dynamic fractal elements. The simple rules of dynamic fractal buildup act like a computer program that generates simple and complex patterns depending on the finite number of its computations - interactions. Initial fundamental elements generate smaller secondary ones. The universe complicates toward the smaller scales with the increase of the number of interacting fundamental elements there. The larger 3D-spiral-vortexes of basic matter move faster inward, oscillate and eject smaller ones, thus creating the finite fundamental elements of the universe in one simple, everything else annihilating, cyclic pattern.

    The principle of parsimony requires not multiply the number of entities in the description of the phenomena. There is no reason why the universe should run uphill if the running downhill firework universe accounts for the many puzzling observations with one entity - one basic matter that self-similarly evolves in the smallest number of dimensions (3D), that allow the creation of the all-explaining, ordered along their origin, finite sources of reality. These sources of unifying interaction evolve in a cyclic mode that annihilates everything else. The non-zero basic matter flux (Fig. 2) indicates that there is something everywhere. The 3D-spiral pattern in which it remains always finite reveals the all-explaining hierarchy of fundamental elements origin. This is also hierarchy of size – the smaller fundamental elements cast from the insides of similar finite larger ones.

**REFERENCES**

1. Mandelbrot, B. B. On fractal geometry, and a few of the mathematical questions it has raised, in: *Proceedings of the International Congress of Mathematicians*, Warszawa, August 16-24, 1983, edited by Z. Ciesielski and C. Olech, PWN - Polish Scientific Publishers - Warszawa and Elsevier Science Publishers, B.V., P.O. Box 1991, 1000 BZ Amsterdam, The Netherlands, Vol. 2, pp. 1661-1675.

2. Savov, E. *Theory of Interaction the Simplest Explanation of Everything.* Geones Books, 2002.
3. Savov, E. P. On the closure of the magnetospheric currents. *Compt. rend. Acad. bulg. Sci.* **44**, No. 5, 29-32 (1991).
4. Laurence, E., A. Bunker, E. Stanway, M. Lacy, R. Ellis and M. Doherty. Spitzer Imaging of i'-drop Galaxies: Old Stars at z~6. *Mon. Not. R. Astron. Soc.*, in press, astro-ph/0502385 (2005).
5. Ouchi, M., K. Shimasaku, M. Akiyama, K. Sekiguchi, H. Furusawa, S. Okamura, N. Kashikawa, M. Iye, T. Kodama, T. Saito, T. Sasaki, C. Simpson, T. Takata, T. Yamada, H. Yamanoi, M. Yoshida and M. Yoshida. The Discovery of Primeval Large-Scale Structures with Forming Clusters at Redshift 6. *Astrophys. J.* **620** L1-L4, astro-ph/0412648 (2005).
6. Elston, R., K. L. Thompson and G. J. Hill. Detection of strong iron emission from quasars at redshift z>3. *Nature* **367**, 250-251 (1994).
7. Maddox, J. Big bang not yet dead but in decline. *Nature* **377**, 99 (1995).
8. Savov, E. The Pattern of Solar Wind-Magnetosphere Interaction and Its Universality. physics/0506075 (2005).
9. Savov, E. P. On the character of solar-terrestrial interactions. *Bulg. Geophys. J.* **19**, No. 4, 57-63 (1993).
10. Manuel O. K. and G. Hwaung. Solar Abundances of the Elements. *Meteoritics* **18**, No. 3 pp. 209-222 (1983).
11. Nodland, B. and J. P. Ralston. Indication of anisotropy in electromagnetic propagation over cosmological distances. *Phys. Rev. Lett.* **78**, 3043-3046 (1997).
12. Konacki, M. An extrasolar giant planet in a close triple-star system. *Nature* **436**, 230-233 (2005).
13. Kukushkin A. B. and V. A. Rantsev-Kartinov. Similarity of skeletal objects in the range $10^{-5}$ cm to $10^{23}$ cm. *Phys. Lett. A* **306**, 175–183 (2002).
14. Kukushkin, A. B. and V. A. Rantsev-Kartinov. Similarity of structuring in the range $10^{-5}$ cm to $10^{23}$ cm hints at a baryonic cold dark skeleton of the Universe. *Bull. Am. Phys. Soc.*, 2002, **47**, November, 2002, astro-ph/0205534 (2002).
15. Dressler, A. The Great Attractor: Do galaxies trace the large-scale mass distribution? *Nature* **350**, 391-397 (1991).
16. Savov, E. P. On the magnetic storm-substorm relationship. *Bulg. Geophys. J.* **24**, Nos. 3/4, 39-49 (1998).
17. Savov, E. Magnetic Storm-substorm Relationship and Some Associated Issues. physics/0501048 (2005).
18. Strogatz, S. *Sync: The Emerging Science of Spontaneous Order.* Theia, 2003.
19. Russell, C. T., R. L. McPherron and R. K. Burton. On the cause of geomagnetic storms. *J. Geophys. Res.* **79**, 1105-1109 (1974).
20. Longair, M. The new astrophysics, in: *The New Physics*, edited by P. Davies, Cambridge University Press, 1989, pp. 94-208.
21. Anderson, J. D., P. A. Laing, E. L. Lau, A. S. Liu, M. M. Nieto and S. G. Turyshev. The study of apparent anomalous, weak, long-range acceleration of Pioneer 10 and 11. *Phys. Rev. D* **65**, 082004/1–50 (2002).
22. Amram, P., C. Mendes de Oliveira, J. Boulesteix and C. Balkowski. The Hα kinematic of the Cartwheel galaxy. *Astron. Astrophys.* **330**, 881-893 (1998).
23. Savov, E. Theory of Interaction Generated from Revealed Universality of Solar Wind Magnetosphere Coupling, Abstract, *5th Understanding Complex Systems Symposium*, University of Illinois at Urbana-Champaign, May 16-19, 2005, http://www.how-why.com/ucs2005/abstracts/Savov.html
24. Arp, H. *Quasars, Redshifts and Controversies.* Interstellar Media, Berkeley, 1987.
25. Arp, H. *Seeing Red: Redshifts, Cosmology and Academic Science.* Aperion, Montreal, 1998.
26. Galianni, P., E. M. Burbidge, H. Arp, V. Junkkarinen, G. Burbidge and S. Zibetti. The Discovery of a High Redshift X-ray Emitting QSO Very Close to the Nucleus of NGC 7319. *Astrophys. J.* **620**, 88-94 (2005).
27. Gutierrez, C. M. and M. Lopez-Corredoira. The Nature of Ultra Luminous X-ray Sources. *Astrophys. J.* **622**, L89-L92 (2005).
28. Arp, H., E. M. Burbidge and G. Burbidge. The double radio source 3C343.1: A galaxy-QSO pair with very different redshifts. *Astron. Astrophys. Lett.* in press. astro-ph/0401007 (2004).
29. Arp, H. *Catalogue of Discordant Redshift Associations (Spiral-bound)*, Aperion, Montreal, 2003.
30. van der Marel, R. P., J. Gerssen, P. Guhathakurta, R. C. Peterson and K. Gebhardt. Hubble Space Telescope Evidence for an Intermediate-Mass Black Hole in the lobular Cluster M15 - I. STIS Spectroscopy and WFPC2 Photometry. *Astron. J.*, **124** (2002) 3255-3270.
31. Gerssen, J., R. P. van der Marel, K. Gebhardt, P. Guhathakurta, R. C. Peterson and C. Pryor. Hubble Space Telescope Evidence for an Intermediate-Mass Black Hole in the Globular Cluster M15 - II. Kinematical Analysis and Dynamical Modeling. *Astron. J.*, **124** (2002) 3270.

32. van der Marel, R., Intermediate-Mass Black Holes in the Universe: A Review of Formation Theories and Observational Constraints. in: *Carnegie Observatories Astrophysics Series, 1: Coevolution of Black Holes and Galaxies*, edited by L. C. Ho (Cambridge: Cambridge Univ. Press) 2003. astro-ph/0302101
33. Van Hilten, D. Palaeomagnetic indications of an increase in the Earth's radius. *Nature* **200**, 1277-1279 (1963).
34. Van Hilten, D. Global expansion and paleomagnetic data. *Tectonophysics* **5**, 191-210 (1968).
35. Meservey, R. Topological inconsistency of continental drift on the present-sized Earth. *Science* **166**, 609-611 (1969).
36. Steward, A. D. Quantitative limits to the paleoradius of the Earth. in: *Expanding Earth Symposium,* edited by S. W. Carey, Sydney, 1981, University of Tasmania, 1983, pp. 305-319.
37. Carey, S. W. *Theories of the Earth and Universe: A History of Dogma in the Earth Sciences.* Stanford University Press, 1988.
38. Owen, H. The Earth Is Expanding and We Don't Know Why. *New Scientist*, November 22, p. 27, (1984).
39. Marchis, F., P. Descamps, D. Hestroffer and J. Berthier. Discovery of the triple asteroidal system 87 Sylvia. *Nature* **436,** 822-824 (2005).
40. Manuel, O. and S. Friberg. Composition of the solar interior: Information from isotope ratios. in *Local and Global Helioseismology: The Present and Future*, Proc. of 2002 SOHO 12 /GONG + 2002, Big Bear Lake, CA USA, edited by H. Lacoste, ESA SP-517, 2003, pp. 345-348.
41. Manuel, O., W. A. Myers, Y. Singh and M. Pleess. The oxygen to carbon ratio in the solar interior: Information from nuclear reaction cross-sections. *J. Fusion Energy* **23**, No. 1, 55-62 (2005).
42. Williams, G. E. Geological constraints on the Precambrian history of Earth's rotation and the Moon's orbit. *Rev. Geophys.* **38**, No.1, 37-59 (2000).
43. Celerier, M. -N. and R. Thieberger. Fractal dimensions of the galaxy distribution varying by steps? astro-ph/0504442 (2005).
44. Joyce, M., F. S. Labini, A. Gabrielli, M. Montuori and L. Pietronero. Basic properties of galaxy clustering in the light of recent results from the Sloan Digital Sky Survey. astro-ph/0501583 (2005).
45. Weidenschilling, S. J., F. Marzari, D. R. Davis and C. Neese. Origin of the Double Asteroid 90 Antiope: A Continuing Puzzle. in *32nd Annual Lunar and Planetary Science Conference*, March 12 -16, 2001, Houston, Texas, abstract no.1890.
46. Michalowski, T., P. Bartszak, F. P. Velichko, A. Kryszczynska, T. Kwiatkowski, S. Breiter, F. Colas, S. Fauvaud, A. Marciniak, J. Michalowski, R. Hirsch, R. Behrend, L. Bernasconi, C. Rinner and S. Charbonnet. Eclipsing binary asteroid 90 Antiope. *Astron. Astrophys.* **423**, 1159-1168 (2004).
47. Zhang, J., X. Song, Y. Li, P. G. Richards, X. Sun and F. Waldhauser. Inner Core Differential Motion Confirmed by Earthquake Waveform Doublets. *Science* **309**, 1357-1360 (2005).
48. Bender, R., J. Kormendy, G. Bower, R. Green, J. Thomas, A. C. Danks, T. Gull, J. B. Hutchings, C. L. Joseph, M. E. Kaiser, T. R. Lauer, C. H. Nelson, D. Richstone, D. Weistrop and B. Woodgate. HST STIS spectroscopy of the triple nucleus of M31: Two nested disks in Keplerian rotation around a supermassive black hole. *Astrophys. J.*, **631**, 280–300 (2005).
49. Manuel, O., S. A. Kamat and M. Mozina. Isotopes Tell Sun's Origin and Operation. in: *Proceedings of the First Crisis in Cosmology Conference,* Monção, Portugal, 23-25 June 2005, Published by the America Institute of Physics, in press. http://web.umr.edu/~om/abstracts2005/IsotopesTellSunsOriginOperation.pdf
50. Rantsev-Kartinov, V. A. Large Scale Self-Similar Skeletal Structure of the Universe. in: *Proceedings of the First Crisis in Cosmology Conference,* Monção, Portugal, 23-25 June 2005, Published by the America Institute of Physics, in press.
51. Kukushkin, A. B. Hypothesis for a Baryonic Cold Skeleton of the Universe as an Implication of Phenomenon of Universal Skeletal Structures. in: *Proceedings of the First Crisis in Cosmology Conference,* Monção, Portugal, 23-25 June 2005, Published by the America Institute of Physics, in press.